\documentclass{aa}  

\usepackage{graphicx}
\usepackage{xcolor}
\usepackage{ulem}
 \usepackage{lscape}

\usepackage{natbib,twoopt}
\usepackage[breaklinks=true, colorlinks=true,linkcolor=blue, citecolor=blue]{hyperref}
\bibpunct{(}{)}{;}{a}{}{,}             
\makeatletter
  \newcommandtwoopt{\citeads}[3][][]{\href{http://adsabs.harvard.edu/abs/#3}%
    {\def\hyper@linkstart##1##2{}%
     \let\hyper@linkend\@empty\citealp[#1][#2]{#3}}}
  \newcommandtwoopt{\citepads}[3][][]{\href{http://adsabs.harvard.edu/abs/#3}%
    {\def\hyper@linkstart##1##2{}%
     \let\hyper@linkend\@empty\citep[#1][#2]{#3}}}
  \newcommandtwoopt{\citetads}[3][][]{\href{http://adsabs.harvard.edu/abs/#3}%
    {\def\hyper@linkstart##1##2{}%
     \let\hyper@linkend\@empty\citet[#1][#2]{#3}}}
  \newcommandtwoopt{\citeyearads}[3][][]%
    {\href{http://adsabs.harvard.edu/abs/#3}
    {\def\hyper@linkstart##1##2{}%
     \let\hyper@linkend\@empty\citeyear[#1][#2]{#3}}}
\makeatother
\defcitealias{das2021}{Paper~I}
\defcitealias{das2024}{Paper~II}
\defcitealias{das2025}{Paper~III}
\usepackage{txfonts}
\begin{document}

   \title{A theoretical framework for BL~Her stars\\ IV. New period-luminosity relations in the Rubin--LSST filters}

   \author{Susmita Das\inst{1,2,3},
           L\'aszl\'o Moln\'ar\inst{1,2,4},
           R\'obert Szab\'o\inst{1,2,4},
           Harinder P. Singh\inst{5},
           Shashi M. Kanbur\inst{6},  
           Anupam Bhardwaj\inst{3},
           Marcella Marconi\inst{7}
           \and
           Radoslaw Smolec \inst{8}
          }

   \institute{Konkoly Observatory, HUN-REN Research Centre for Astronomy and Earth Sciences, Konkoly-Thege Mikl\'os \'ut 15-17, H-1121, Budapest, Hungary\\
              \email{susmita.das@csfk.org}
        \and
             CSFK, MTA Centre of Excellence, Budapest, Konkoly Thege Miklós út 15-17., H-1121, Hungary
         \and
            Inter-University Center for Astronomy and Astrophysics (IUCAA), Post Bag 4, Ganeshkhind, Pune 411007, India
        \and
            ELTE E\"otv\"os Lor\'and University, Institute of Physics and Astronomy, 1117, P\'azm\'any P\'eter s\'et\'any 1/A, Budapest, Hungary
         \and
             Department of Physics \& Astrophysics, University of Delhi, Delhi 110007, India
         \and
            Department of Physics, State University of New York Oswego, Oswego, NY 13126, USA
        \and
             INAF-Osservatorio Astronomico di Capodimonte, Salita Moiariello 16, 80131, Naples, Italy
        \and
            Nicolaus Copernicus Astronomical Center, Polish Academy of Sciences, Bartycka 18, PL-00-716 Warsaw, Poland
         }
\authorrunning{S. Das et al.}
   \date{Received 2 October 2024; accepted 11 January 2025}

 
  \abstract
   {The upcoming Rubin--LSST is expected to revolutionise the field of classical pulsators with well-sampled multi-epoch photometric data in multiple wavelengths. Type~II Cepheids (T2Cs) exhibit weak or negligible metallicity dependence on period-luminosity ($PL$) relations and may potentially be used as an alternative to classical Cepheids for extragalactic distance estimations, when used together with RR~Lyraes and the tip of the red giant branch. It is therefore crucial to study an updated theoretical pulsation scenario of BL~Herculis stars (BL~Her; the shortest period T2Cs) in the corresponding Rubin--LSST photometric system.}
   {We present new theoretical light curves in the Rubin--LSST filters for a fine grid of BL~Her models computed using \textsc{mesa-rsp}. We also derive new theoretical $PL$ and period-Wesenheit ($PW$) relations in the Rubin--LSST filters with the goal to study the effect of convection parameters and metallicity on these relations.}
  {The grid of BL~Her models was computed using the non-linear radial stellar pulsation tool \textsc{mesa-rsp} with the input stellar parameters:\ metallicity ($-2.0\; \mathrm{dex} \leq \mathrm{[Fe/H]} \leq 0.0\; \mathrm{dex}$), stellar mass ($0.5M_{\odot}-0.8M_{\odot}$), stellar luminosity ($50L_{\odot}-300L_{\odot}$), and effective temperature (across the full extent of the instability strip; in steps of 50K) and using four sets of convection parameters. Bolometric correction tables from MIST were used to transform the theoretical bolometric light curves of the BL~Her models into the Rubin--LSST $ugrizy$ filters.}
   {The $PL$ relations of the BL~Her models exhibit steeper slopes but smaller dispersion with increasing wavelengths in the Rubin--LSST filters. The $PL$ and $PW$ slopes for the complete set of BL~Her models computed with radiative cooling (sets B and D) are statistically similar across the $grizy$ filters. The BL~Her models exhibit weak or negligible effect of metallicity on the $PL$ relations for wavelengths longer than the $g$ filter for both the cases of the complete set of models as well as the low-mass models. However, we find significant effect of metallicity on the $PL$ relation in the $u$ filter. Strong metallicity effects are observed in the $PWZ$ relations involving the $u$ filter and are found to have significant contribution from the high-metallicity BL~Her models. Due to negligible metallicity effect for relations involving the Wesenheit indices $W(i,g-i)$, $W(z,i-z)$ and $W(y,g-y)$, we recommend these filter combinations for BL~Her stars when observed with the Rubin--LSST to be used as reliable standard candles.}
   {}

   \keywords{hydrodynamics- methods: numerical- stars: oscillations (including pulsations)- stars: Population II- stars: variables: Cepheids- stars: low-mass}

   \maketitle

\section{Introduction}

BL~Herculis (BL~Her) stars belong to the class of Type~II Cepheids (T2Cs) which lie intermediate to RR~Lyraes and classical Cepheids on the instability strip of the Hertzsprung-Russell Diagram \citep[HRD,][]{gingold1985, wallerstein2002, bono2024}. BL~Her stars are typically considered to be low-mass stars \citep{bono1997a, caputo1998, bono2020} with pulsation periods between 1 and 4 days \citep{soszynski2018}. It is important to note that both the lower and the upper limits on pulsation period are not strict. The pulsation period of 1 day separates RR~Lyraes and T2Cs \citep{soszynski2008, soszynski2014}; however, the RR~Lyrae--T2C separation is a long-standing problem and is now recommended to be separated on a more general evolution-dependent threshold \citep{braga2020, bono2024}. An in-depth analysis of the light curve morphology of these classical pulsators may also be useful in this regard \citep[see, for example, Fig.~9 of][]{smolec2018}. On the other hand, the upper limit of the pulsation period of BL~Her stars is 4~days in the Magellanic Clouds \citep{soszynski2018} but 5~days in the Galactic bulge \citep{soszynski2017}, depending on the stellar environment. However, in this study, we consider the conventional classification of pulsation periods $1 \leq P (\textrm {days}) \leq 4$ for BL~Her stars.

BL~Her stars (and T2Cs) obey well-defined period-luminosity ($PL$) relations \citep{matsunaga2006, groenewegen2008, matsunaga2009, ciechanowska2010, matsunaga2011, ripepi2015, bhardwaj2017b, bhardwaj2017c, groenewegen2017b, braga2018, bhardwaj2022,ngeow2022,sicignano2024}, similar to RR~Lyrae stars (in near-infrared) and classical Cepheids, and are therefore useful distance estimators \citep[see reviews,][]{beaton2018b, bhardwaj2020a, bono2024}. However, BL~Her stars seem to have an advantage over the other classical pulsators with their weak or negligible effect of metallicity on the $PL$ relations, as demonstrated by the empirical studies mentioned above and complemented by theoretical studies \citep{criscienzo2007, das2021, das2024}. Recently, several new T2Cs have been detected towards the Galactic center \citep{braga2019} as well as from the VIrac VAriable Classification Ensemble (VIVACE) catalogue \citep{molnar2022} of the Vista Variables in the V\'ia L\'actea (VVV) infrared survey of the Galactic bar/bulge and southern disc. In addition, \citet{wielgorski2024} recently determined projection factors for eight nearby BL~Her stars and calibrated the semi-geometric Baade-Wesselink method \citep{baade1926, becker1940, wesselink1947} for measuring distances using BL~Her stars.

This paper is in preparation for the upcoming Vera C.\ Rubin Observatory Legacy Survey of Space and Time (Rubin--LSST)\footnote{\url{https://rubinobservatory.org/}} which is expected to revolutionise the field of stellar astrophysics (and in particular, classical pulsators). The new ground-based telescope is expected to detect stars at limiting magnitudes of around 25~mag{\footnote{For more details, see \url{https://smtn-002.lsst.io/}}}, thereby producing unprecedented astronomical data of the deep universe. The Rubin--LSST will observe the southern sky in six filters $u_\textrm{LSST}, g_\textrm{LSST}, r_\textrm{LSST}, i_\textrm{LSST}, z_\textrm{LSST}$ and $y_\textrm{LSST}$ (hereafter $ugrizy$) and will allow us to obtain well-sampled multi-epoch photometry of the classical pulsators, among other astronomical data. We therefore provide multi-band theoretical $PL$ relations of BL~Her models in the Rubin--LSST filters for a future comparison with the observations, as was recently done by \citet{marconi2022} for RR~Lyrae models.

We use the fine grid of convective BL~Her models first presented in \citet{das2021} (hereafter \citetalias{das2021}) and subsequently used in \citet{das2024} and \citet{das2025} (hereafter \citetalias{das2024} and \citetalias{das2025}, respectively). The grid of models was computed using the state-of-the-art 1D non-linear Radial Stellar Pulsation (\textsc{rsp}) tool within the \textit{Modules for Experiments in Stellar Astrophysics} \citep[\textsc{mesa},][]{paxton2011,paxton2013,paxton2015,paxton2018,paxton2019,jermyn2023} software suite for a wide range of input parameters: metallicity ($-2.0\; \mathrm{dex} \leq \mathrm{[Fe/H]} \leq 0.0\; \mathrm{dex}$), stellar mass (0.5--0.8\,$M_{\odot}$), stellar luminosity (50--300\,$L_{\odot}$), and effective temperature (full extent of the instability strip; in steps of 50K) and using four sets of convection parameters as outlined in \citet{paxton2019}. We presented new theoretical $PL$ relations for BL~Her models in the Johnson-Cousins-Glass bands ($UBVRIJHKLL'M$) in \citetalias{das2021} and in the $Gaia$ passbands ($G$, $G_{BP}$, and $G_{RP}$) in \citetalias{das2024} and compared with the empirical relations at mean light from \citet{matsunaga2006, matsunaga2009, matsunaga2011, bhardwaj2017c, groenewegen2017b, ripepi2023}. One of the most interesting results was the weak or negligible metallicity effect on the theoretical $PL$ relations in wavelengths longer than $B$ band as was observed in the empirical relations. In \citetalias{das2025}, we carried out a robust light curve optimisation technique to compare stellar pulsation BL~Her models computed using \textsc{mesa-rsp} with observed light curves. We applied the technique to 58 BL~Her stars in the Large Magellanic Cloud from $Gaia$~DR3 over their entire pulsation cycle and not just at mean light.

As an extension to \citetalias{das2021} and \citetalias{das2024}, here we compute light curves for the fine grid of BL~Her models in the Rubin--LSST filters ($ugrizy$), derive new multi-band theoretical $PL$ relations and thereby study the effects of metallicity and convection parameters. The structure of this paper is as follows: in Section~\ref{sec:data}, we present the theoretical light curves of the BL~Her models encompassing a wide range of stellar mass, stellar luminosity, metallicities and effective temperature across the four sets of convection parameters in the Rubin--LSST filters. In Section~\ref{sec:PL}, we derive the theoretical $PL$ relations in the $ugrizy$ filters and thereby study the effect of metallicity and convection parameters on these relations. A comparison of the $PL$ relations of BL~Her models with those from RR~Lyrae models \citep{marconi2022} in the Rubin--LSST filters is carried out in Section~\ref{sec:RRL}. Lastly, the results of this study are summarised in Section~\ref{sec:results}.

\section{Theoretical light curves in the Rubin--LSST filters}
\label{sec:data}

\begin{figure*}
\centering
\includegraphics[scale = 0.9]{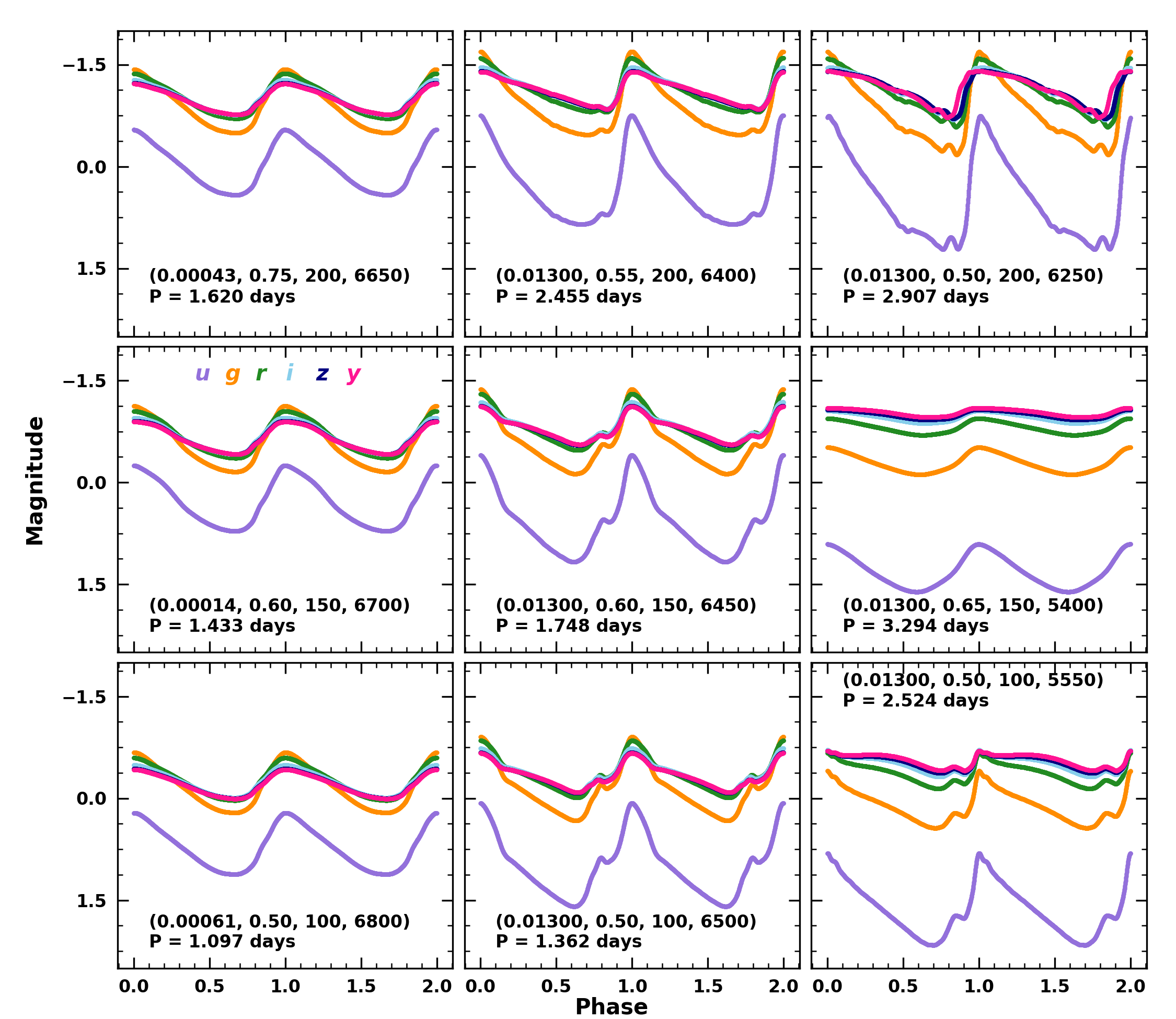}
\caption{Theoretical light curves of a few BL~Her models in the Rubin-LSST filters computed using convection parameter set~A. The input stellar parameters of the corresponding models are included in the format ($Z, M/M_{\odot}, L/L_{\odot}, T_{\rm eff}$) in each sub-plot. The increasing stellar luminosity ($L/L_{\odot}$) are plotted from the bottom to the top panels while the increasing effective temperature ($T_{\rm eff}$) are displayed from the right to the left panels.}
\label{fig:LC}
\end{figure*}

\subsection{Stellar pulsation models}
We use the grid of non-linear radial stellar pulsation BL~Her models computed using \textsc{mesa-rsp} \citep{smolec2008, paxton2019} in \citetalias{das2021}. In brief, the grid of BL~Her models encompasses over: (i) seven metallicities, $Z = 0.00014, 0.00043, 0.00061, 0.00135, 0.00424, 0.00834, 0.013$ (see Table~\ref{tab:composition} for more details) (ii) stellar masses $M/M_{\odot} \in \left\langle 0.5, 0.8\right\rangle$ with a step of $0.05\, M_{\odot}$ (iii) stellar luminosities $L/L_{\odot} \in \left\langle 50, 300\right\rangle$\footnote{The low mass models $M/M_{\odot} \in [0.5, 0.6]$ were computed with stellar luminosity $L/L_{\odot} \in \left\langle 50, 200\right\rangle$ only.} with a step of $50\,L_\odot$ (iv) effective temperatures $T_{\rm{eff}} \in \left\langle4000, 8000\right\rangle$ K with a step of 50 K. Each grid with these $ZXMLT_{\rm eff}$ input stellar parameters was computed using the four sets of convection parameters (sets A, B, C and D), each with increasing complexity as outlined in Table~4 of \citet{paxton2019}. The non-linear computations for the grid of models with the above-mentioned input stellar parameters and convective efficiencies were carried for 4000 pulsation cycles each and thereby checked for the conditions of full-amplitude stable pulsation\footnote{The fractional growth of the kinetic energy per pulsation period, $\Gamma$ does not vary by more than 0.01, the amplitude of radius variation, $\Delta R$ by more than 0.01~$R_{\odot}$, and the pulsation period, $P$ by more than 0.01~d over the last 100 cycles of the completed integration (see Fig.~2 of \citetalias{das2021} for a detailed pictorial representation).} and single periodicity. Following the above checks and considering only those models with non-linear pulsation periods between 1--4 days to be BL~Her models \citep{soszynski2018}, the number of BL~Her models accepted in each convection set are as follows: 3266~(set~A), 2260~(set~B), 2632~(set~C) and 2122~(set~D).

It is important to highlight here that our grid of models includes stellar masses higher than what is typically considered for BL~Her stars \citepalias[for more discussion, see][]{das2021,das2024}. BL~Her stars are predicted to be low-mass stars and in a double-shell (hydrogen and helium) burning phase \citep{salaris2005}, as they evolve off the Zero-Age Horizontal Branch (ZAHB) and approach their Asymptotic Giant Branch (AGB) track \citep{gingold1985, bono1997a, caputo1998, bono2020, braga2020, bono2024}. The higher stellar mass (M\,$> 0.6\,M_{\odot}$) and lower metallicity ($Z=0.00014$) models may also be considered typical of evolved RR~Lyrae stars. However, as mentioned in \citetalias{das2021} and \citetalias{das2024}, we explore the possibility of higher stellar mass models following a few theoretical studies on BL~Her stars \citep{marconi2007, smolec2012b} and also to accommodate the differences in the uncertainties between different evolutionary codes. In \citetalias{das2025}, we indeed found the majority of observed BL~Hers in the LMC to match well with the lower stellar mass models ($M \leq 0.65\,M_{\odot}$), with just a few preferring the higher stellar mass models. In light of this, for a reliable comparison with future observations from the Rubin--LSST, we derive the $PL$ relations and carry out the theoretical analysis twice: for the complete set of BL~Her models with $M/M_{\odot} \in [0.5, 0.8]$ and also for the low-mass models only with $M/M_{\odot} \in [0.5, 0.6]$.

\begin{table*}
\caption{Light curve parameters of the BL~Her models used in this analysis computed using \textsc{mesa-rsp}. The columns provide the chemical composition ($ZX$), stellar mass ($\frac{M}{M_{\odot}}$), stellar luminosity ($\frac{L}{L_{\odot}}$), effective temperature ($T_{\rm eff}$), pulsation period ($P$), the mean magnitudes and the peak-to-peak amplitudes in each of the Rubin-LSST filters, and the convection parameter set used.}
\centering
\scalebox{0.7}{
\begin{tabular}{c c c c c c c c c c c c c c c c c c c}
\hline\hline
$Z$ & $X$ & $M$ & $L$ & $T_{\rm eff}$ &  $P$ & $\langle u_{\rm LSST} \rangle$ & $A_u$ & $\langle g_{\rm LSST} \rangle$ & $A_g$ &$\langle r_{\rm LSST} \rangle$ & $A_r$ &$\langle i_{\rm LSST} \rangle$ & $A_i$ &$\langle z_{\rm LSST} \rangle$ & $A_z$ &$\langle y_{\rm LSST} \rangle$ & $A_y$ & Convection\\
& & $(M_{\odot})$ & $(L_{\odot})$ & (K) & (days) & (mag) & (mag) & (mag) & (mag) & (mag) & (mag) & (mag) & (mag) & (mag) & (mag) & (mag) & (mag) & set\\
[0.5ex]
\hline \hline
0.01300&	0.71847&	0.50&	50&	5900&	1.0875&	1.969&	0.158&	0.712&	0.101&	0.359&	0.063&	0.278&	0.049&	0.277&	0.043&	0.267&	0.039&	A\\
0.01300&	0.71847&	0.50&	50&	5950&	1.0523&	1.934&	0.235&	0.697&	0.154&	0.359&	0.095&	0.284&	0.076&	0.286&	0.067&	0.278&	0.063&	A\\
...&    ...&    ...&    ...&    ...&    ...&    ...&    ...&    ...&    ...&    ...&    ...& ...&    ...&    ...&    ...&    ...&   ...\\
\\
0.01300&	0.71847&	0.50&	50&	5900&	1.0919&	1.980&	0.258&	0.716&	0.165&	0.360&	0.103&	0.277&	0.082&	0.275&	0.072&	0.264&	0.068&	B\\
0.01300&	0.71847&	0.50&	50&	5950&	1.0558&	1.946&	0.343&	0.702&	0.230&	0.360&	0.142&	0.283&	0.117&	0.284&	0.105&	0.275&	0.100&	B\\
...&    ...&    ...&    ...&    ...&    ...&    ...&    ...&    ...&    ...&    ...&    ...& ...&    ...&    ...&    ...&    ...&   ...\\
\\
0.01300&	0.71847&	0.50&	50&	5650&	1.3319&	2.276&	0.023&	0.836&	0.014&	0.374&	0.009&	0.244&	0.007&	0.217&	0.006&	0.192&	0.005&	C\\
0.01300&	0.71847&	0.50&	50&	5700&	1.2774&	2.223&	0.354&	0.816&	0.224&	0.372&	0.153&	0.250&	0.124&	0.227&	0.109&	0.206&	0.102&	C\\
...&    ...&    ...&    ...&    ...&    ...&    ...&    ...&    ...&    ...&    ...&    ...& ...&    ...&    ...&    ...&    ...&   ...\\
\\
0.01300&	0.71847&	0.50&	50&	5700&	1.2798&	2.224&	0.319&	0.816&	0.198&	0.372&	0.132&	0.250&	0.106&	0.227&	0.092&	0.205&	0.086&	D\\
0.01300&	0.71847&	0.50&	50&	5750&	1.2290&	2.167&	0.371&	0.793&	0.236&	0.368&	0.163&	0.255&	0.134&	0.236&	0.119&	0.217&	0.112&	D\\
...&    ...&    ...&    ...&    ...&    ...&    ...&    ...&    ...&    ...&    ...&    ...& ...&    ...&    ...&    ...&    ...&   ...\\
\\
\hline
\end{tabular}}
\tablefoot{\small 
        This table is available entirely in electronic form at the CDS.}
\label{tab:allmodels}
\end{table*}

\subsection{Processing the model data}

The non-linear computations of the models using \textsc{mesa-rsp} result in bolometric light curves, which can then be transformed into light curves in different passbands  \citep[for more details, see,][]{paxton2018}. While \textsc{mesa} includes a few pre-computed bolometric correction (BC) tables, it does not yet include BC tables to transform the bolometric light curves into the different Rubin--LSST filters. As in \citetalias{das2024}, we therefore include user-provided BC tables defined as a function of the stellar photosphere in terms of $T_{\rm eff}$\,(K), $\log (g ({\rm cm \; s^{-2}}))$, and metallicity [M/H]. We use BC tables obtained from the MIST\footnote{\url{https://waps.cfa.harvard.edu/MIST/index.html}} (\textsc{mesa} Isochrones \& Stellar Tracks) packaged model grids which were computed using 1D atmosphere models \citep[based on ATLAS12/SYNTHE;][]{kurucz1970, kurucz1993}, thereby transforming the theoretical bolometric light curves of our grid of BL~Her models into the Rubin--LSST $ugrizy$ filters. A few examples of the theoretical light curves of BL~Her models in the Rubin-LSST filters computed using convection parameter set~A are presented in Fig.~\ref{fig:LC}. The theoretical BL~Her light curves in the Rubin-LSST filters across the different convection parameter sets are available upon request.

Instead of using a simple mean magnitude to obtain the $PL$ relations in the different filters, we use the mean magnitudes obtained by fitting the theoretical light curves in the Rubin--LSST filters with the Fourier sine series \citep[see, for example,][]{deb2009, das2020} of the form:
\begin{equation}
m(x) = m_0 + \sum_{k=1}^{N}A_k \sin(2 \pi kx+\phi_k),
\label{eq:fourier}
\end{equation}

\noindent where $x$ is the instantaneous pulsation phase, $m_0$ is the mean magnitude, and $N$ is the order of the fit ($N = 20$). We also obtain the peak-to-peak amplitude $A$ of the light curve as the difference in the minimum and the maximum of the light variations:
\begin{equation}
    A_{\lambda} = (M_{\lambda})_{\rm min} - (M_{\lambda})_{\rm max},
\end{equation}
where $(M_{\lambda})_{\rm min}$ and $(M_{\lambda})_{\rm max}$ are the minimum and maximum magnitudes obtained from the Fourier fits in the $\lambda$ band, respectively. The input stellar parameters as well as the computed light curve parameters of the BL~Her models in the different Rubin--LSST filters obtained from the Fourier fitting and used in this analysis are provided in Table~\ref{tab:allmodels}. 

From both Fig.~\ref{fig:LC} and Table~\ref{tab:allmodels}, we find that the peak-to-peak amplitude of the light curve of a particular BL~Her model decreases with increasing wavelength as we go from the $u$ to the $y$ filters. This is well-known and is expected because the effective temperature of a BL~Her star (similar to RR~Lyraes and classical Cepheids) causes the Planck function to peak at visible wavelengths with the temperature dependence of visible flux scaling as $R^2T_\textrm{eff}^4$. The temperature dependence of infrared flux, on the other hand, lies on the Rayleigh-Jeans tail of the Planck function and scales as $R^2T_\textrm{eff}^{1.6}$ \citep{jameson1986, smith2015, das2018}. This is probed further in the Bailey diagram (peak-to-peak amplitude versus period) as a function of stellar mass, chemical composition and wavelength displayed in Fig.~\ref{fig:bailey}. As discussed earlier, amplitude decreases with increasing wavelength. However, a higher stellar mass model exhibits a slightly larger amplitude at a particular wavelength. We also find a significantly larger amplitude for the higher metallicity model, corresponding to the same stellar mass and luminosity; this is more pronounced at the shorter wavelengths ($u_{\rm LSST}$ and $g_{\rm LSST}$). Similar results hold for all four sets of convection parameters.

\section{Period--Luminosity relations}
\label{sec:PL}

The theoretical period-luminosity ($PL$) relations of the BL~Her models in the Rubin--LSST filters are derived using the  mathematical form:
\begin{equation}
M_\lambda = a\log(P)+b,
\end{equation}
where $M_\lambda$ refers to the mean magnitudes obtained from the Fourier-fitted theoretical light curves in a given band, $\lambda$. Note that a few empirical and theoretical studies use intensity-averaged magnitudes which may result in slightly different $PL$ relations, especially at the shorter wavelengths.

In addition, we also use Wesenheit indices \citep{madore1982} which offer the advantage of being minimally affected by uncertainties arising from extinction corrections. The Wesenheit magnitude combinations adopted for this work come from \citet{marconi2022} and are defined as follows:
\begin{equation}
\begin{aligned}
W_{ug} &:=  W(g,u-g) &= g-3.100(u-g); \\
W_{ur} &:=  W(r,u-r) &= r-1.258(u-r); \\
W_{gr} &:=  W(r,g-r) &= r-2.796(g-r); \\
W_{gi} &:=  W(i,g-i) &= i-1.287(g-i); \\
W_{iz} &:=  W(z,i-z) &= z-3.204(i-z); \\
W_{gy} &:=  W(y,g-y) &= y-0.560(g-y), 
\end{aligned}
\label{eq:W}
\end{equation}
where $ugrizy$ are the absolute mean magnitudes in the respective Rubin--LSST filters. The color term coefficients result from the extinction law by \citet{cardelli1989}, assuming $R_V=3.1$. We thereby also derive theoretical multi-filter period-Wesenheit ($PW$) relations of the BL~Her models of the mathematical form:
\begin{equation}
W_{\lambda_2, \lambda_1} = a\log(P)+b,
\end{equation}
where $\lambda_1 > \lambda_2$ and the multi-filter Wesenheit indices $W_{\lambda_2, \lambda_1}$ are obtained from Eq.~\ref{eq:W}.

\begin{table*}
\caption{Comparison of the slopes of the $PL$ relations for BL~Her stars of the mathematical form $M_\lambda=a\log(P)+b$. The theoretical relations are derived for the cases of the complete set of models and for the low mass models. $N$ is the total number of models. |$T$| represents the observed value of the $t$-statistic, and $p(t)$ gives the probability of acceptance of the null hypothesis (equal slopes). The bold-faced entries indicate that the null hypothesis of the equivalent $PL$ slopes can be rejected.}
\centering
\scalebox{0.9}{
\begin{tabular}{c c c c c c c c c}
\hline\hline
Band & Source & $a$ & $b$ & $\sigma$ & $N$ & \multicolumn{3}{c}{(|$T$|, $p(t)$) w.r.t.}\\
(LSST) & & & & & & Set A & Set B & Set C\\
\hline \hline
\multicolumn{9}{c}{Complete set of models ($0.5-0.8M_{\odot}$)}\\
\hline
$u$ & $\rm{Z_{all}}$ (Set A) &-0.834$\pm$0.044&0.815$\pm$0.015&0.39&3266& ... & ... & ... \\
$u$ & $\rm{Z_{all}}$ (Set B) &-0.589$\pm$0.046&0.848$\pm$0.015&0.352&2260& \textbf{(3.843,0.0}& ... & ... \\
$u$ & $\rm{Z_{all}}$ (Set C) &-0.475$\pm$0.05&0.944$\pm$0.017&0.418&2632& \textbf{(5.412,0.0)} & (1.69,0.091) & ... \\
$u$ & $\rm{Z_{all}}$ (Set D) &-0.405$\pm$0.052&0.952$\pm$0.018&0.4&2122& \textbf{(6.308,0.0)} & \textbf{(2.65,0.008)} & (0.961,0.336)\\
\hline
$g$ & $\rm{Z_{all}}$ (Set A) &-1.423$\pm$0.035&-0.089$\pm$0.012&0.312&3266& ... & ... & ... \\
$g$ & $\rm{Z_{all}}$ (Set B) &-1.164$\pm$0.036&-0.077$\pm$0.012&0.278&2260& \textbf{(5.121,0.0)}& ... & ... \\
$g$ & $\rm{Z_{all}}$ (Set C) &-1.241$\pm$0.038&0.033$\pm$0.013&0.319&2632& \textbf{(3.53,0.0)} & (1.45,0.147) & ... \\
$g$ & $\rm{Z_{all}}$ (Set D) &-1.095$\pm$0.039&0.023$\pm$0.013&0.302&2122& \textbf{(6.241,0.0)} & (1.297,0.195) & \textbf{(2.669,0.008)}\\
\hline
$r$ & $\rm{Z_{all}}$ (Set A) &-1.829$\pm$0.028&-0.23$\pm$0.01&0.253&3266& ... & ... & ... \\
$r$ & $\rm{Z_{all}}$ (Set B) &-1.602$\pm$0.03&-0.232$\pm$0.01&0.231&2260& \textbf{(5.466,0.0)}& ... & ... \\
$r$ & $\rm{Z_{all}}$ (Set C) &-1.705$\pm$0.031&-0.146$\pm$0.011&0.26&2632& \textbf{(2.947,0.003)} & \textbf{(2.387,0.017)} & ... \\
$r$ & $\rm{Z_{all}}$ (Set D) &-1.563$\pm$0.032&-0.159$\pm$0.011&0.249&2122& \textbf{(6.199,0.0)} & (0.888,0.375) & \textbf{(3.193,0.001)}\\
\hline
$i$ & $\rm{Z_{all}}$ (Set A) &-2.01$\pm$0.025&-0.238$\pm$0.009&0.225&3266& ... & ... & ... \\
$i$ & $\rm{Z_{all}}$ (Set B) &-1.804$\pm$0.027&-0.245$\pm$0.009&0.209&2260& \textbf{(5.509,0.0)}& ... & ... \\
$i$ & $\rm{Z_{all}}$ (Set C) &-1.902$\pm$0.028&-0.175$\pm$0.01&0.232&2632& \textbf{(2.871,0.004)} & \textbf{(2.514,0.012)} & ... \\
$i$ & $\rm{Z_{all}}$ (Set D) &-1.771$\pm$0.029&-0.188$\pm$0.01&0.224&2122& \textbf{(6.189,0.0)} & (0.826,0.409) & \textbf{(3.262,0.001)}\\
\hline
$z$ & $\rm{Z_{all}}$ (Set A) &-2.099$\pm$0.024&-0.217$\pm$0.008&0.214&3266& ... & ... & ... \\
$z$ & $\rm{Z_{all}}$ (Set B) &-1.903$\pm$0.026&-0.226$\pm$0.009&0.199&2260& \textbf{(5.507,0.0)}& ... & ... \\
$z$ & $\rm{Z_{all}}$ (Set C) &-2.001$\pm$0.026&-0.162$\pm$0.009&0.219&2632& \textbf{(2.751,0.006)} & \textbf{(2.647,0.008)} & ... \\
$z$ & $\rm{Z_{all}}$ (Set D) &-1.876$\pm$0.028&-0.174$\pm$0.009&0.213&2122& \textbf{(6.093,0.0)} & (0.718,0.473) & \textbf{(3.294,0.001)}\\
\hline
$y$ & $\rm{Z_{all}}$ (Set A) &-2.15$\pm$0.023&-0.21$\pm$0.008&0.209&3266& ... & ... & ... \\
$y$ & $\rm{Z_{all}}$ (Set B) &-1.957$\pm$0.026&-0.22$\pm$0.008&0.196&2260& \textbf{(5.537,0.0)}& ... & ... \\
$y$ & $\rm{Z_{all}}$ (Set C) &-2.061$\pm$0.025&-0.158$\pm$0.009&0.214&2632& \textbf{(2.565,0.01)} & \textbf{(2.869,0.004)} & ... \\
$y$ & $\rm{Z_{all}}$ (Set D) &-1.936$\pm$0.027&-0.17$\pm$0.009&0.208&2122& \textbf{(5.979,0.0)} & (0.569,0.569) & \textbf{(3.367,0.001)}\\
\hline
\multicolumn{9}{c}{Low-mass models only ($0.5-0.6M_{\odot}$)}\\
\hline
$u$ & $\rm{Z_{all}}$ (Set A) &-0.235$\pm$0.071&0.858$\pm$0.023&0.351&1050& ... & ... & ... \\
$u$ & $\rm{Z_{all}}$ (Set B) &-0.198$\pm$0.067&0.958$\pm$0.022&0.3&707& (0.38,0.704)& ... & ... \\
$u$ & $\rm{Z_{all}}$ (Set C) &-0.252$\pm$0.081&1.122$\pm$0.028&0.396&856& (0.155,0.877) & (0.512,0.609) & ... \\
$u$ & $\rm{Z_{all}}$ (Set D) &-0.388$\pm$0.08&1.169$\pm$0.028&0.384&711& (1.428,0.153) & (1.811,0.07) & (1.196,0.232)\\
\hline
$g$ & $\rm{Z_{all}}$ (Set A) &-0.92$\pm$0.055&-0.036$\pm$0.018&0.274&1050& ... & ... & ... \\
$g$ & $\rm{Z_{all}}$ (Set B) &-0.798$\pm$0.049&0.025$\pm$0.016&0.218&707& (1.666,0.096)& ... & ... \\
$g$ & $\rm{Z_{all}}$ (Set C) &-1.043$\pm$0.059&0.193$\pm$0.02&0.29&856& (1.513,0.13) & \textbf{(3.193,0.001)} & ... \\
$g$ & $\rm{Z_{all}}$ (Set D) &-1.034$\pm$0.057&0.208$\pm$0.02&0.274&711& (1.427,0.154) & \textbf{(3.136,0.002)} & (0.107,0.915)\\
\hline
$r$ & $\rm{Z_{all}}$ (Set A) &-1.433$\pm$0.043&-0.155$\pm$0.014&0.213&1050& ... & ... & ... \\
$r$ & $\rm{Z_{all}}$ (Set B) &-1.304$\pm$0.039&-0.12$\pm$0.013&0.172&707& \textbf{(2.237,0.025)}& ... & ... \\
$r$ & $\rm{Z_{all}}$ (Set C) &-1.544$\pm$0.046&0.011$\pm$0.016&0.226&856& (1.765,0.078) & \textbf{(4.001,0.0)} & ... \\
$r$ & $\rm{Z_{all}}$ (Set D) &-1.502$\pm$0.045&0.015$\pm$0.016&0.213&711& (1.109,0.267) & \textbf{(3.355,0.001)} & (0.662,0.508) \\
\hline
$i$ & $\rm{Z_{all}}$ (Set A) &-1.664$\pm$0.037&-0.152$\pm$0.012&0.182&1050& ... & ... & ... \\
$i$ & $\rm{Z_{all}}$ (Set B) &-1.544$\pm$0.033&-0.126$\pm$0.011&0.149&707& \textbf{(2.413,0.016)}& ... & ... \\
$i$ & $\rm{Z_{all}}$ (Set C) &-1.761$\pm$0.039&-0.018$\pm$0.014&0.194&856& (1.791,0.073) & \textbf{(4.185,0.0)} & ... \\
$i$ & $\rm{Z_{all}}$ (Set D) &-1.718$\pm$0.038&-0.015$\pm$0.013&0.184&711& (1.011,0.312) & \textbf{(3.408,0.001)} & (0.778,0.437)\\
\hline
$z$ & $\rm{Z_{all}}$ (Set A) &-1.773$\pm$0.034&-0.127$\pm$0.011&0.169&1050& ... & ... & ... \\
$z$ & $\rm{Z_{all}}$ (Set B) &-1.659$\pm$0.031&-0.104$\pm$0.01&0.139&707& \textbf{(2.47,0.014)}& ... & ... \\
$z$ & $\rm{Z_{all}}$ (Set C) &-1.868$\pm$0.037&-0.006$\pm$0.013&0.179&856& (1.902,0.057) & \textbf{(4.357,0.0)} & ... \\
$z$ & $\rm{Z_{all}}$ (Set D) &-1.826$\pm$0.036&-0.003$\pm$0.012&0.171&711& (1.068,0.286) & \textbf{(3.521,0.0)} & (0.829,0.407)\\
\hline
$y$ & $\rm{Z_{all}}$ (Set A) &-1.833$\pm$0.033&-0.118$\pm$0.011&0.164&1050& ... & ... & ... \\
$y$ & $\rm{Z_{all}}$ (Set B) &-1.718$\pm$0.03&-0.098$\pm$0.01&0.135&707& \textbf{(2.56,0.011)}& ... & ... \\
$y$ & $\rm{Z_{all}}$ (Set C) &-1.932$\pm$0.035&-0.002$\pm$0.012&0.173&856& \textbf{(2.041,0.041)} & \textbf{(4.591,0.0)} & ... \\
$y$ & $\rm{Z_{all}}$ (Set D) &-1.885$\pm$0.035&-0.0$\pm$0.012&0.165&711& (1.09,0.276) & \textbf{(3.632,0.0)} & (0.942,0.346)\\
\hline
\end{tabular}}
\label{tab:PL}
\end{table*}

\begin{table*}
\caption{Comparison of the slopes of the $PW$ relations for BL~Her stars of the mathematical form $W_\lambda=a\log(P)+b$. The theoretical relations are derived for the cases of the complete set of models and for the low mass models. $N$ is the total number of models. |$T$| represents the observed value of the $t$-statistic, and $p(t)$ gives the probability of acceptance of the null hypothesis (equal slopes). The bold-faced entries indicate that the null hypothesis of the equivalent $PL$ slopes can be rejected.}
\centering
\scalebox{1}{
\begin{tabular}{c c c c c c c c c}
\hline\hline
Band & Source & $a$ & $b$ & $\sigma$ & $N$ & \multicolumn{3}{c}{(|$T$|, $p(t)$) w.r.t.}\\
(LSST) & & & & & & Set A & Set B & Set C\\
\hline \hline
\multicolumn{9}{c}{Complete set of models ($0.5-0.8M_{\odot}$)}\\
\hline
$W_{ug}$ & $\rm{Z_{all}}$ (Set A) &-3.249$\pm$0.053&-2.892$\pm$0.018&0.469&3266& ... & ... & ... \\
$W_{ug}$ & $\rm{Z_{all}}$ (Set B) &-2.946$\pm$0.064&-2.942$\pm$0.021&0.488&2260 & \textbf{(3.653,0.0)}& ... & ... \\
$W_{ug}$ & $\rm{Z_{all}}$ (Set C) &-3.615$\pm$0.063&-2.792$\pm$0.022&0.529&2632 & \textbf{(4.45,0.0)} & \textbf{(7.444,0.0)}  & ... \\
$W_{ug}$ & $\rm{Z_{all}}$ (Set D) &-3.232$\pm$0.07&-2.857$\pm$0.024&0.541&2122 &(0.194,0.846) & \textbf{(3.009,0.003)} & \textbf{(4.057,0.0)}\\
\hline
$W_{ur}$ & $\rm{Z_{all}}$ (Set A) &-3.081$\pm$0.031&-1.545$\pm$0.011&0.279&3266& ... & ... & ...\\
$W_{ur}$ & $\rm{Z_{all}}$ (Set B) &-2.876$\pm$0.038&-1.589$\pm$0.012&0.29&2260 & \textbf{(4.167,0.0)}& ... & ... \\
$W_{ur}$ & $\rm{Z_{all}}$ (Set C) &-3.254$\pm$0.037&-1.517$\pm$0.013&0.307&2632 & \textbf{(3.575,0.0)} & \textbf{(7.155,0.0)}  & ... \\
$W_{ur}$ & $\rm{Z_{all}}$ (Set D) &-3.018$\pm$0.041&-1.557$\pm$0.014&0.316&2122 &(1.218,0.223) & \textbf{(2.554,0.011)} & \textbf{(4.282,0.0)}\\
\hline
$W_{gr}$ & $\rm{Z_{all}}$ (Set A) &-2.965$\pm$0.019&-0.624$\pm$0.007&0.172&3266& ... & ... & ... \\
$W_{gr}$ & $\rm{Z_{all}}$ (Set B) &-2.826$\pm$0.023&-0.665$\pm$0.008&0.178&2260 & \textbf{(4.59,0.0)}& ... & ... \\
$W_{gr}$ & $\rm{Z_{all}}$ (Set C) &-3.005$\pm$0.021&-0.645$\pm$0.007&0.179&2632 & (1.386,0.166) & \textbf{(5.671,0.0)}  & ... \\
$W_{gr}$ & $\rm{Z_{all}}$ (Set D) &-2.87$\pm$0.024&-0.668$\pm$0.008&0.185&2122 &\textbf{(3.054,0.002)} & (1.341,0.18) & \textbf{(4.179,0.0)}\\
\hline
$W_{gi}$ & $\rm{Z_{all}}$ (Set A) &-2.765$\pm$0.017&-0.431$\pm$0.006&0.152&3266& ... & ... & ... \\
$W_{gi}$ & $\rm{Z_{all}}$ (Set B) &-2.628$\pm$0.02&-0.462$\pm$0.007&0.154&2260 & \textbf{(5.144,0.0)}& ... & ... \\
$W_{gi}$ & $\rm{Z_{all}}$ (Set C) &-2.754$\pm$0.019&-0.442$\pm$0.006&0.156&2632 & (0.439,0.661) & \textbf{(4.555,0.0}  & ... \\
$W_{gi}$ & $\rm{Z_{all}}$ (Set D) &-2.642$\pm$0.021&-0.459$\pm$0.007&0.16&2122 &\textbf{(4.548,0.0)} & (0.482,0.63) & \textbf{(3.991,0.0)}\\
\hline
$W_{iz}$ & $\rm{Z_{all}}$ (Set A) &-2.384$\pm$0.02&-0.149$\pm$0.007&0.181&3266& ... & ... & ... \\
$W_{iz}$ & $\rm{Z_{all}}$ (Set B) &-2.219$\pm$0.023&-0.164$\pm$0.007&0.174&2260 & \textbf{(5.383,0.0)}& ... & ... \\
$W_{iz}$ & $\rm{Z_{all}}$ (Set C) &-2.318$\pm$0.022&-0.12$\pm$0.008&0.184&2632 & \textbf{(2.207,0.027)} & \textbf{(3.115,0.002)}  & ... \\
$W_{iz}$ & $\rm{Z_{all}}$ (Set D) &-2.21$\pm$0.024&-0.131$\pm$0.008&0.182&2122 &\textbf{(5.574,0.0)} & (0.282,0.778) & \textbf{(3.344,0.001)}\\
\hline
$W_{gy}$ & $\rm{Z_{all}}$ (Set A) &-2.557$\pm$0.018&-0.278$\pm$0.006&0.164&3266& ... & ... & ... \\
$W_{gy}$ & $\rm{Z_{all}}$ (Set B) &-2.401$\pm$0.021&-0.3$\pm$0.007&0.161&2260 & \textbf{(5.538,0.0)}& ... & ... \\
$W_{gy}$ & $\rm{Z_{all}}$ (Set C) &-2.521$\pm$0.02&-0.264$\pm$0.007&0.167&2632 & (1.341,0.18) & \textbf{(4.095,0.0)}  & ... \\
$W_{gy}$ & $\rm{Z_{all}}$ (Set D) &-2.407$\pm$0.022&-0.279$\pm$0.007&0.168&2122 &\textbf{(5.243,0.0)} & (0.191,0.849) & \textbf{(3.835,0.0)}\\
\hline
\multicolumn{9}{c}{Low-mass models only ($0.5-0.6M_{\odot}$)}\\
\hline
$W_{ug}$ & $\rm{Z_{all}}$ (Set A) &-3.046$\pm$0.095&-2.808$\pm$0.031&0.472&1050& ... & ... & ... \\
$W_{ug}$ & $\rm{Z_{all}}$ (Set B) &-2.658$\pm$0.111&-2.866$\pm$0.036&0.492&707 & \textbf{(2.662,0.008)}& ... & ... \\
$W_{ug}$ & $\rm{Z_{all}}$ (Set C) &-3.496$\pm$0.111&-2.687$\pm$0.038&0.544&856 & \textbf{(3.079,0.002)} & \textbf{(5.35,0.0)}  & ... \\
$W_{ug}$ & $\rm{Z_{all}}$ (Set D) &-3.037$\pm$0.115&-2.771$\pm$0.04&0.552&711& (0.055,0.956) & \textbf{(2.378,0.018)} & \textbf{(2.862,0.004)}\\
\hline
$W_{ur}$ & $\rm{Z_{all}}$ (Set A) &-2.94$\pm$0.054&-1.43$\pm$0.017&0.266&1050& ... & ... & ... \\
$W_{ur}$ & $\rm{Z_{all}}$ (Set B) &-2.695$\pm$0.062&-1.475$\pm$0.02&0.276&707 & \textbf{(2.995,0.003)}& ... & ... \\
$W_{ur}$ & $\rm{Z_{all}}$ (Set C) &-3.17$\pm$0.062&-1.387$\pm$0.021&0.302&856 & \textbf{(2.817,0.005)} & \textbf{(5.437,0.0)}  & ... \\
$W_{ur}$ & $\rm{Z_{all}}$ (Set D) &-2.903$\pm$0.064&-1.436$\pm$0.022&0.306&711 &(0.445,0.657) & \textbf{(2.338,0.02)} & \textbf{(3.007,0.003)}\\
\hline
$W_{gr}$ & $\rm{Z_{all}}$ (Set A) &-2.866$\pm$0.027&-0.488$\pm$0.009&0.134&1050& ... & ... & ... \\
$W_{gr}$ & $\rm{Z_{all}}$ (Set B) &-2.718$\pm$0.031&-0.524$\pm$0.01&0.137&707 & \textbf{(3.606,0.0)}& ... & ... \\
$W_{gr}$ & $\rm{Z_{all}}$ (Set C) &-2.946$\pm$0.029&-0.499$\pm$0.01&0.145&856 & \textbf{(1.992,0.046)} & \textbf{(5.332,0.0)}  & ... \\
$W_{gr}$ & $\rm{Z_{all}}$ (Set D) &-2.809$\pm$0.031&-0.523$\pm$0.011&0.147&711 &(1.385,0.166) & \textbf{(2.098,0.036)} & \textbf{(3.203,0.001)}\\
\hline
$W_{gi}$ & $\rm{Z_{all}}$ (Set A) &-2.621$\pm$0.02&-0.301$\pm$0.007&0.099&1050& ... & ... & ... \\
$W_{gi}$ & $\rm{Z_{all}}$ (Set B) &-2.505$\pm$0.022&-0.32$\pm$0.007&0.097&707 & \textbf{(3.931,0.0)}& ... & ... \\
$W_{gi}$ & $\rm{Z_{all}}$ (Set C) &-2.684$\pm$0.021&-0.291$\pm$0.007&0.105&856 & \textbf{(2.168,0.03)} & \textbf{(5.894,0.0)}  & ... \\
$W_{gi}$ & $\rm{Z_{all}}$ (Set D) &-2.598$\pm$0.022&-0.302$\pm$0.008&0.105&711& (0.779,0.436) & \textbf{(3.001,0.003)} & \textbf{(2.823,0.005)}\\
\hline
$W_{iz}$ & $\rm{Z_{all}}$ (Set A) &-2.122$\pm$0.026&-0.045$\pm$0.009&0.131&1050& ... & ... & ... \\
$W_{iz}$ & $\rm{Z_{all}}$ (Set B) &-2.026$\pm$0.025&-0.033$\pm$0.008&0.111&707 & \textbf{(2.655,0.008)}& ... & ... \\
$W_{iz}$ & $\rm{Z_{all}}$ (Set C) &-2.212$\pm$0.028&0.035$\pm$0.009&0.135&856 & \textbf{(2.364,0.018)} & \textbf{(5.015,0.0)}  & ... \\
$W_{iz}$ & $\rm{Z_{all}}$ (Set D) &-2.172$\pm$0.028&0.037$\pm$0.01&0.132&711 &(1.298,0.194) & \textbf{(3.924,0.0)} & (1.042,0.297)\\
\hline
$W_{gy}$ & $\rm{Z_{all}}$ (Set A) &-2.345$\pm$0.022&-0.164$\pm$0.007&0.112&1050& ... & ... & ... \\
$W_{gy}$ & $\rm{Z_{all}}$ (Set B) &-2.234$\pm$0.022&-0.166$\pm$0.007&0.099&707 & \textbf{(3.494,0.0)}& ... & ... \\
$W_{gy}$ & $\rm{Z_{all}}$ (Set C) &-2.43$\pm$0.024&-0.112$\pm$0.008&0.116&856 & \textbf{(2.612,0.009)} & \textbf{(6.026,0.0)}  & ... \\
$W_{gy}$ & $\rm{Z_{all}}$ (Set D) &-2.362$\pm$0.024&-0.117$\pm$0.008&0.114&711 &(0.54,0.59) & \textbf{(3.926,0.0)} & \textbf{(2.008,0.045)}\\
\hline
\end{tabular}}
\label{tab:PW}
\end{table*}

\begin{table}[ht!]
\caption{$PLZ$ relations for BL~Her models of the mathematical form $M_\lambda=\alpha+\beta\log(P)+\gamma\mathrm{[Fe/H]}$ for different wavelengths using different convective parameter sets.}
\centering
\scalebox{0.8}{
\begin{tabular}{c c c c c c}
\hline\hline
Band & $\alpha$ & $\beta$ & $\gamma$ & $\sigma$ & $N$\\
\hline \hline
\multicolumn{6}{c}{Complete set of models ($0.5-0.8M_{\odot}$)}\\
\hline
\multicolumn{6}{c}{Convection set A}\\
\hline
$u$ & 1.057 $\pm$ 0.017 & -0.983 $\pm$ 0.041 & 0.223 $\pm$ 0.009 & 0.359 & 3266\\
$g$ & -0.062 $\pm$ 0.015 & -1.44 $\pm$ 0.035 & 0.025 $\pm$ 0.008 & 0.311 & 3266\\
$r$ & -0.251 $\pm$ 0.012 & -1.816 $\pm$ 0.029 & -0.019 $\pm$ 0.007 & 0.253 & 3266\\
$i$ & -0.254 $\pm$ 0.011 & -2.0 $\pm$ 0.026 & -0.014 $\pm$ 0.006 & 0.225 & 3266\\
$z$ & -0.223 $\pm$ 0.01 & -2.095 $\pm$ 0.024 & -0.005 $\pm$ 0.006 & 0.214 & 3266\\
$y$ & -0.218 $\pm$ 0.01 & -2.145 $\pm$ 0.024 & -0.008 $\pm$ 0.005 & 0.209 & 3266\\
\hline
\multicolumn{6}{c}{Convection set B}\\
\hline
$u$ & 1.073 $\pm$ 0.017 & -0.693 $\pm$ 0.042 & 0.221 $\pm$ 0.01 & 0.318 & 2260\\
$g$ & -0.061 $\pm$ 0.015 & -1.171 $\pm$ 0.037 & 0.015 $\pm$ 0.009 & 0.277 & 2260\\
$r$ & -0.261 $\pm$ 0.012 & -1.588 $\pm$ 0.03 & -0.029 $\pm$ 0.007 & 0.23 & 2260\\
$i$ & -0.268 $\pm$ 0.011 & -1.794 $\pm$ 0.027 & -0.022 $\pm$ 0.006 & 0.208 & 2260\\
$z$ & -0.239 $\pm$ 0.011 & -1.897 $\pm$ 0.026 & -0.013 $\pm$ 0.006 & 0.199 & 2260\\
$y$ & -0.236 $\pm$ 0.01 & -1.95 $\pm$ 0.026 & -0.016 $\pm$ 0.006 & 0.195 & 2260\\
\hline
\multicolumn{6}{c}{Convection set C}\\
\hline
$u$ & 1.218 $\pm$ 0.02 & -0.685 $\pm$ 0.046 & 0.255 $\pm$ 0.011 & 0.381 & 2632\\
$g$ & 0.059 $\pm$ 0.016 & -1.261 $\pm$ 0.039 & 0.024 $\pm$ 0.009 & 0.318 & 2632\\
$r$ & -0.172 $\pm$ 0.013 & -1.685 $\pm$ 0.032 & -0.024 $\pm$ 0.007 & 0.259 & 2632\\
$i$ & -0.194 $\pm$ 0.012 & -1.887 $\pm$ 0.028 & -0.018 $\pm$ 0.007 & 0.231 & 2632\\
$z$ & -0.171 $\pm$ 0.011 & -1.994 $\pm$ 0.027 & -0.009 $\pm$ 0.006 & 0.219 & 2632\\
$y$ & -0.169 $\pm$ 0.011 & -2.052 $\pm$ 0.026 & -0.011 $\pm$ 0.006 & 0.214 & 2632\\
\hline
\multicolumn{6}{c}{Convection set D}\\
\hline
$u$ & 1.213 $\pm$ 0.02 & -0.553 $\pm$ 0.047 & 0.256 $\pm$ 0.011 & 0.36 & 2122\\
$g$ & 0.046 $\pm$ 0.017 & -1.108 $\pm$ 0.04 & 0.023 $\pm$ 0.01 & 0.302 & 2122\\
$r$ & -0.185 $\pm$ 0.014 & -1.548 $\pm$ 0.032 & -0.026 $\pm$ 0.008 & 0.248 & 2122\\
$i$ & -0.207 $\pm$ 0.012 & -1.76 $\pm$ 0.029 & -0.019 $\pm$ 0.007 & 0.223 & 2122\\
$z$ & -0.184 $\pm$ 0.012 & -1.87 $\pm$ 0.028 & -0.01 $\pm$ 0.007 & 0.213 & 2122\\
$y$ & -0.183 $\pm$ 0.011 & -1.929 $\pm$ 0.027 & -0.012 $\pm$ 0.007 & 0.208 & 2122\\
\hline
\multicolumn{6}{c}{Low-mass models only ($0.5-0.6M_{\odot}$)}\\
\hline
\multicolumn{6}{c}{Convection set A}\\
\hline
$u$ & 1.081 $\pm$ 0.027 & -0.394 $\pm$ 0.066 & 0.206 $\pm$ 0.015 & 0.322 & 1050\\
$g$ & -0.03 $\pm$ 0.023 & -0.925 $\pm$ 0.056 & 0.006 $\pm$ 0.012 & 0.274 & 1050\\
$r$ & -0.195 $\pm$ 0.017 & -1.405 $\pm$ 0.043 & -0.036 $\pm$ 0.01 & 0.212 & 1050\\
$i$ & -0.185 $\pm$ 0.015 & -1.641 $\pm$ 0.037 & -0.03 $\pm$ 0.008 & 0.181 & 1050\\
$z$ & -0.149 $\pm$ 0.014 & -1.757 $\pm$ 0.035 & -0.021 $\pm$ 0.008 & 0.169 & 1050\\
$y$ & -0.143 $\pm$ 0.013 & -1.816 $\pm$ 0.033 & -0.023 $\pm$ 0.007 & 0.163 & 1050\\
\hline
\multicolumn{6}{c}{Convection set B}\\
\hline
$u$ & 1.174 $\pm$ 0.024 & -0.279 $\pm$ 0.058 & 0.224 $\pm$ 0.014 & 0.258 & 707\\
$g$ & 0.036 $\pm$ 0.02 & -0.802 $\pm$ 0.049 & 0.011 $\pm$ 0.012 & 0.217 & 707\\
$r$ & -0.153 $\pm$ 0.016 & -1.291 $\pm$ 0.038 & -0.034 $\pm$ 0.009 & 0.171 & 707\\
$i$ & -0.153 $\pm$ 0.013 & -1.534 $\pm$ 0.033 & -0.028 $\pm$ 0.008 & 0.147 & 707\\
$z$ & -0.122 $\pm$ 0.013 & -1.652 $\pm$ 0.031 & -0.019 $\pm$ 0.008 & 0.138 & 707\\
$y$ & -0.118 $\pm$ 0.012 & -1.711 $\pm$ 0.03 & -0.021 $\pm$ 0.007 & 0.134 & 707\\
\hline
\multicolumn{6}{c}{Convection set C}\\
\hline
$u$ & 1.391 $\pm$ 0.031 & -0.453 $\pm$ 0.073 & 0.262 $\pm$ 0.018 & 0.354 & 856\\
$g$ & 0.216 $\pm$ 0.025 & -1.06 $\pm$ 0.06 & 0.022 $\pm$ 0.015 & 0.29 & 856\\
$r$ & -0.02 $\pm$ 0.02 & -1.521 $\pm$ 0.047 & -0.03 $\pm$ 0.011 & 0.225 & 856\\
$i$ & -0.043 $\pm$ 0.017 & -1.742 $\pm$ 0.04 & -0.024 $\pm$ 0.01 & 0.193 & 856\\
$z$ & -0.021 $\pm$ 0.016 & -1.856 $\pm$ 0.037 & -0.015 $\pm$ 0.009 & 0.179 & 856\\
$y$ & -0.02 $\pm$ 0.015 & -1.918 $\pm$ 0.036 & -0.018 $\pm$ 0.009 & 0.172 & 856\\
\hline
\multicolumn{6}{c}{Convection set D}\\
\hline
$u$ & 1.417 $\pm$ 0.03 & -0.476 $\pm$ 0.071 & 0.271 $\pm$ 0.019 & 0.337 & 711\\
$g$ & 0.231 $\pm$ 0.024 & -1.042 $\pm$ 0.057 & 0.025 $\pm$ 0.015 & 0.274 & 711\\
$r$ & -0.009 $\pm$ 0.019 & -1.493 $\pm$ 0.045 & -0.027 $\pm$ 0.012 & 0.213 & 711\\
$i$ & -0.035 $\pm$ 0.016 & -1.711 $\pm$ 0.038 & -0.021 $\pm$ 0.01 & 0.183 & 711\\
$z$ & -0.014 $\pm$ 0.015 & -1.822 $\pm$ 0.036 & -0.013 $\pm$ 0.009 & 0.171 & 711\\
$y$ & -0.015 $\pm$ 0.015 & -1.88 $\pm$ 0.035 & -0.015 $\pm$ 0.009 & 0.165 & 711\\
\hline
\end{tabular}}
\label{tab:PLZ}
\end{table}

\begin{table}[ht!]
\caption{$PLZ$ relations for BL~Her models of the mathematical form $M_\lambda=\alpha+\beta\log(P)+\gamma\mathrm{[Fe/H]}$ in the low- and the high-metallicity regimes for different wavelengths using different convective parameter sets.}
\centering
\scalebox{0.8}{
\begin{tabular}{c c c c c c}
\hline\hline
Band & $\alpha$ & $\beta$ & $\gamma$ & $\sigma$ & $N$\\
\hline \hline
\multicolumn{6}{c}{Low-metallicity regime ($Z=0.00135, 0.00061, 0.00043, 0.00014$)}\\
\hline
\multicolumn{6}{c}{Convection set A}\\
\hline
$u$ & 0.923 $\pm$ 0.037 & -1.182 $\pm$ 0.053 & 0.098 $\pm$ 0.022 & 0.331 & 1734\\
$g$ & -0.103 $\pm$ 0.034 & -1.454 $\pm$ 0.05 & -0.005 $\pm$ 0.021 & 0.31 & 1734\\
$r$ & -0.264 $\pm$ 0.028 & -1.796 $\pm$ 0.041 & -0.025 $\pm$ 0.017 & 0.255 & 1734\\
$i$ & -0.267 $\pm$ 0.025 & -1.978 $\pm$ 0.036 & -0.019 $\pm$ 0.015 & 0.227 & 1734\\
$z$ & -0.24 $\pm$ 0.024 & -2.074 $\pm$ 0.035 & -0.013 $\pm$ 0.014 & 0.215 & 1734\\
$y$ & -0.234 $\pm$ 0.023 & -2.122 $\pm$ 0.034 & -0.014 $\pm$ 0.014 & 0.211 & 1734\\
\hline
\multicolumn{6}{c}{Convection set B}\\
\hline
$u$ & 0.964 $\pm$ 0.036 & -0.918 $\pm$ 0.05 & 0.11 $\pm$ 0.022 & 0.277 & 1217\\
$g$ & -0.074 $\pm$ 0.034 & -1.205 $\pm$ 0.047 & 0.001 $\pm$ 0.021 & 0.263 & 1217\\
$r$ & -0.25 $\pm$ 0.029 & -1.579 $\pm$ 0.04 & -0.02 $\pm$ 0.018 & 0.223 & 1217\\
$i$ & -0.26 $\pm$ 0.026 & -1.779 $\pm$ 0.036 & -0.014 $\pm$ 0.016 & 0.202 & 1217\\
$z$ & -0.236 $\pm$ 0.025 & -1.88 $\pm$ 0.035 & -0.008 $\pm$ 0.015 & 0.194 & 1217\\
$y$ & -0.232 $\pm$ 0.025 & -1.93 $\pm$ 0.034 & -0.009 $\pm$ 0.015 & 0.191 & 1217\\
\hline
\multicolumn{6}{c}{Convection set C}\\
\hline
$u$ & 1.044 $\pm$ 0.041 & -0.795 $\pm$ 0.057 & 0.121 $\pm$ 0.025 & 0.32 & 1292\\
$g$ & -0.005 $\pm$ 0.038 & -1.159 $\pm$ 0.053 & 0.0 $\pm$ 0.023 & 0.296 & 1292\\
$r$ & -0.202 $\pm$ 0.032 & -1.574 $\pm$ 0.044 & -0.024 $\pm$ 0.019 & 0.247 & 1292\\
$i$ & -0.223 $\pm$ 0.028 & -1.79 $\pm$ 0.039 & -0.019 $\pm$ 0.017 & 0.222 & 1292\\
$z$ & -0.202 $\pm$ 0.027 & -1.903 $\pm$ 0.038 & -0.013 $\pm$ 0.016 & 0.212 & 1292\\
$y$ & -0.199 $\pm$ 0.026 & -1.963 $\pm$ 0.037 & -0.014 $\pm$ 0.016 & 0.207 & 1292\\
\hline
\multicolumn{6}{c}{Convection set D}\\
\hline
$u$ & 1.055 $\pm$ 0.039 & -0.747 $\pm$ 0.051 & 0.117 $\pm$ 0.024 & 0.284 & 1094\\
$g$ & -0.001 $\pm$ 0.037 & -1.087 $\pm$ 0.048 & -0.004 $\pm$ 0.023 & 0.268 & 1094\\
$r$ & -0.2 $\pm$ 0.031 & -1.493 $\pm$ 0.041 & -0.025 $\pm$ 0.019 & 0.229 & 1094\\
$i$ & -0.222 $\pm$ 0.029 & -1.706 $\pm$ 0.037 & -0.018 $\pm$ 0.018 & 0.21 & 1094\\
$z$ & -0.202 $\pm$ 0.028 & -1.815 $\pm$ 0.036 & -0.011 $\pm$ 0.017 & 0.201 & 1094\\
$y$ & -0.199 $\pm$ 0.027 & -1.872 $\pm$ 0.035 & -0.012 $\pm$ 0.017 & 0.198 & 1094\\
\hline
\multicolumn{6}{c}{High-metallicity regime ($Z=0.01300, 0.00834, 0.00424$)}\\
\hline
\multicolumn{6}{c}{Convection set A}\\
\hline
$u$ & 1.056 $\pm$ 0.026 & -0.822 $\pm$ 0.062 & 0.424 $\pm$ 0.048 & 0.381 & 1532\\
$g$ & -0.049 $\pm$ 0.021 & -1.436 $\pm$ 0.051 & 0.083 $\pm$ 0.04 & 0.313 & 1532\\
$r$ & -0.239 $\pm$ 0.017 & -1.839 $\pm$ 0.041 & 0.001 $\pm$ 0.032 & 0.25 & 1532\\
$i$ & -0.241 $\pm$ 0.015 & -2.025 $\pm$ 0.036 & 0.005 $\pm$ 0.028 & 0.222 & 1532\\
$z$ & -0.209 $\pm$ 0.014 & -2.12 $\pm$ 0.034 & 0.017 $\pm$ 0.027 & 0.211 & 1532\\
$y$ & -0.205 $\pm$ 0.014 & -2.171 $\pm$ 0.033 & 0.013 $\pm$ 0.026 & 0.206 & 1532\\
\hline
\multicolumn{6}{c}{Convection set B}\\
\hline
$u$ & 1.037 $\pm$ 0.027 & -0.446 $\pm$ 0.069 & 0.36 $\pm$ 0.053 & 0.351 & 1043\\
$g$ & -0.071 $\pm$ 0.022 & -1.134 $\pm$ 0.057 & 0.016 $\pm$ 0.044 & 0.293 & 1043\\
$r$ & -0.264 $\pm$ 0.018 & -1.597 $\pm$ 0.047 & -0.056 $\pm$ 0.036 & 0.239 & 1043\\
$i$ & -0.269 $\pm$ 0.016 & -1.81 $\pm$ 0.042 & -0.046 $\pm$ 0.033 & 0.214 & 1043\\
$z$ & -0.238 $\pm$ 0.016 & -1.916 $\pm$ 0.04 & -0.032 $\pm$ 0.031 & 0.205 & 1043\\
$y$ & -0.234 $\pm$ 0.015 & -1.972 $\pm$ 0.039 & -0.034 $\pm$ 0.03 & 0.201 & 1043\\
\hline
\multicolumn{6}{c}{Convection set C}\\
\hline
$u$ & 1.241 $\pm$ 0.031 & -0.623 $\pm$ 0.072 & 0.434 $\pm$ 0.058 & 0.427 & 1340\\
$g$ & 0.099 $\pm$ 0.024 & -1.357 $\pm$ 0.057 & 0.052 $\pm$ 0.046 & 0.337 & 1340\\
$r$ & -0.139 $\pm$ 0.019 & -1.784 $\pm$ 0.045 & -0.031 $\pm$ 0.037 & 0.269 & 1340\\
$i$ & -0.166 $\pm$ 0.017 & -1.974 $\pm$ 0.04 & -0.023 $\pm$ 0.033 & 0.239 & 1340\\
$z$ & -0.143 $\pm$ 0.016 & -2.076 $\pm$ 0.038 & -0.009 $\pm$ 0.031 & 0.226 & 1340\\
$y$ & -0.142 $\pm$ 0.016 & -2.132 $\pm$ 0.037 & -0.011 $\pm$ 0.03 & 0.219 & 1340\\
\hline
\multicolumn{6}{c}{Convection set D}\\
\hline
$u$ & 1.182 $\pm$ 0.033 & -0.349 $\pm$ 0.081 & 0.37 $\pm$ 0.065 & 0.42 & 1028\\
$g$ & 0.053 $\pm$ 0.026 & -1.135 $\pm$ 0.065 & -0.005 $\pm$ 0.052 & 0.333 & 1028\\
$r$ & -0.174 $\pm$ 0.021 & -1.61 $\pm$ 0.052 & -0.075 $\pm$ 0.041 & 0.266 & 1028\\
$i$ & -0.194 $\pm$ 0.019 & -1.823 $\pm$ 0.046 & -0.062 $\pm$ 0.037 & 0.236 & 1028\\
$z$ & -0.169 $\pm$ 0.018 & -1.934 $\pm$ 0.043 & -0.046 $\pm$ 0.035 & 0.223 & 1028\\
$y$ & -0.168 $\pm$ 0.017 & -1.995 $\pm$ 0.042 & -0.047 $\pm$ 0.034 & 0.217 & 1028\\
\hline
\end{tabular}}
\label{tab:PLZ_metallicityregime}
\end{table}

\begin{table}[ht!]
\caption{$PWZ$ relations for BL~Her models of the mathematical form $W_{\lambda_2, \lambda_1}=\alpha+\beta\log(P)+\gamma\mathrm{[Fe/H]}$ for different wavelengths using different convective parameter sets.}
\centering
\scalebox{0.8}{
\begin{tabular}{c c c c c c}
\hline\hline
Band & $\alpha$ & $\beta$ & $\gamma$ & $\sigma$ & $N$\\
\hline \hline
\multicolumn{6}{c}{Complete set of models ($0.5-0.8M_{\odot}$)}\\
\hline
\multicolumn{6}{c}{Convection set A}\\
\hline
$W_{ug}$ & -3.529 $\pm$ 0.012 & -2.856 $\pm$ 0.028 & -0.588 $\pm$ 0.006 & 0.244 & 3266\\
$W_{ur}$ & -1.896 $\pm$ 0.008 & -2.864 $\pm$ 0.019 & -0.324 $\pm$ 0.004 & 0.17 & 3266\\
$W_{gr}$ & -0.78 $\pm$ 0.007 & -2.869 $\pm$ 0.016 & -0.144 $\pm$ 0.004 & 0.142 & 3266\\
$W_{gi}$ & -0.501 $\pm$ 0.007 & -2.722 $\pm$ 0.017 & -0.065 $\pm$ 0.004 & 0.146 & 3266\\
$W_{iz}$ & -0.123 $\pm$ 0.009 & -2.4 $\pm$ 0.021 & 0.024 $\pm$ 0.005 & 0.18 & 3266\\
$W_{gy}$ & -0.306 $\pm$ 0.008 & -2.54 $\pm$ 0.019 & -0.026 $\pm$ 0.004 & 0.163 & 3266\\
\hline
\multicolumn{6}{c}{Convection set B}\\
\hline
$W_{ug}$ & -3.576 $\pm$ 0.013 & -2.653 $\pm$ 0.032 & -0.621 $\pm$ 0.007 & 0.241 & 2260\\
$W_{ur}$ & -1.939 $\pm$ 0.009 & -2.714 $\pm$ 0.023 & -0.342 $\pm$ 0.005 & 0.172 & 2260\\
$W_{gr}$ & -0.82 $\pm$ 0.008 & -2.754 $\pm$ 0.019 & -0.152 $\pm$ 0.004 & 0.144 & 2260\\
$W_{gi}$ & -0.535 $\pm$ 0.008 & -2.595 $\pm$ 0.019 & -0.071 $\pm$ 0.005 & 0.147 & 2260\\
$W_{iz}$ & -0.147 $\pm$ 0.009 & -2.227 $\pm$ 0.023 & 0.017 $\pm$ 0.005 & 0.173 & 2260\\
$W_{gy}$ & -0.334 $\pm$ 0.009 & -2.386 $\pm$ 0.021 & -0.033 $\pm$ 0.005 & 0.16 & 2260\\
\hline
\multicolumn{6}{c}{Convection set C}\\
\hline
$W_{ug}$ & -3.533 $\pm$ 0.013 & -3.044 $\pm$ 0.03 & -0.691 $\pm$ 0.007 & 0.25 & 2632\\
$W_{ur}$ & -1.919 $\pm$ 0.009 & -2.943 $\pm$ 0.021 & -0.376 $\pm$ 0.005 & 0.174 & 2632\\
$W_{gr}$ & -0.816 $\pm$ 0.007 & -2.873 $\pm$ 0.017 & -0.16 $\pm$ 0.004 & 0.143 & 2632\\
$W_{gi}$ & -0.52 $\pm$ 0.008 & -2.694 $\pm$ 0.018 & -0.073 $\pm$ 0.004 & 0.148 & 2632\\
$W_{iz}$ & -0.096 $\pm$ 0.009 & -2.336 $\pm$ 0.022 & 0.022 $\pm$ 0.005 & 0.183 & 2632\\
$W_{gy}$ & -0.297 $\pm$ 0.009 & -2.495 $\pm$ 0.02 & -0.031 $\pm$ 0.005 & 0.166 & 2632\\
\hline
\multicolumn{6}{c}{Convection set D}\\
\hline
$W_{ug}$ & -3.572 $\pm$ 0.014 & -2.829 $\pm$ 0.033 & -0.7 $\pm$ 0.008 & 0.256 & 2122\\
$W_{ur}$ & -1.944 $\pm$ 0.01 & -2.8 $\pm$ 0.024 & -0.379 $\pm$ 0.006 & 0.181 & 2122\\
$W_{gr}$ & -0.832 $\pm$ 0.008 & -2.778 $\pm$ 0.02 & -0.16 $\pm$ 0.005 & 0.149 & 2122\\
$W_{gi}$ & -0.533 $\pm$ 0.008 & -2.6 $\pm$ 0.02 & -0.073 $\pm$ 0.005 & 0.152 & 2122\\
$W_{iz}$ & -0.11 $\pm$ 0.01 & -2.222 $\pm$ 0.024 & 0.021 $\pm$ 0.006 & 0.181 & 2122\\
$W_{gy}$ & -0.311 $\pm$ 0.009 & -2.389 $\pm$ 0.022 & -0.031 $\pm$ 0.005 & 0.167 & 2122\\
\hline
\multicolumn{6}{c}{Low-mass models only ($0.5-0.6M_{\odot}$)}\\
\hline
\multicolumn{6}{c}{Convection set A}\\
\hline
$W_{ug}$ & -3.471 $\pm$ 0.018 & -2.572 $\pm$ 0.046 & -0.612 $\pm$ 0.01 & 0.224 & 1050\\
$W_{ur}$ & -1.8 $\pm$ 0.011 & -2.676 $\pm$ 0.027 & -0.341 $\pm$ 0.006 & 0.13 & 1050\\
$W_{gr}$ & -0.657 $\pm$ 0.007 & -2.745 $\pm$ 0.017 & -0.156 $\pm$ 0.004 & 0.082 & 1050\\
$W_{gi}$ & -0.385 $\pm$ 0.007 & -2.561 $\pm$ 0.017 & -0.077 $\pm$ 0.004 & 0.085 & 1050\\
$W_{iz}$ & -0.034 $\pm$ 0.011 & -2.13 $\pm$ 0.027 & 0.01 $\pm$ 0.006 & 0.131 & 1050\\
$W_{gy}$ & -0.207 $\pm$ 0.009 & -2.314 $\pm$ 0.022 & -0.039 $\pm$ 0.005 & 0.108 & 1050\\
\hline
\multicolumn{6}{c}{Convection set B}\\
\hline
$W_{ug}$ & -3.492 $\pm$ 0.019 & -2.424 $\pm$ 0.048 & -0.648 $\pm$ 0.012 & 0.213 & 707\\
$W_{ur}$ & -1.821 $\pm$ 0.012 & -2.566 $\pm$ 0.029 & -0.359 $\pm$ 0.007 & 0.127 & 707\\
$W_{gr}$ & -0.679 $\pm$ 0.007 & -2.66 $\pm$ 0.019 & -0.161 $\pm$ 0.005 & 0.082 & 707\\
$W_{gi}$ & -0.396 $\pm$ 0.007 & -2.476 $\pm$ 0.018 & -0.078 $\pm$ 0.004 & 0.08 & 707\\
$W_{iz}$ & -0.023 $\pm$ 0.01 & -2.03 $\pm$ 0.025 & 0.011 $\pm$ 0.006 & 0.11 & 707\\
$W_{gy}$ & -0.205 $\pm$ 0.009 & -2.22 $\pm$ 0.022 & -0.04 $\pm$ 0.005 & 0.095 & 707\\
\hline
\multicolumn{6}{c}{Convection set C}\\
\hline
$W_{ug}$ & -3.427 $\pm$ 0.021 & -2.941 $\pm$ 0.05 & -0.723 $\pm$ 0.012 & 0.24 & 856\\
$W_{ur}$ & -1.794 $\pm$ 0.012 & -2.865 $\pm$ 0.029 & -0.397 $\pm$ 0.007 & 0.139 & 856\\
$W_{gr}$ & -0.678 $\pm$ 0.007 & -2.812 $\pm$ 0.017 & -0.174 $\pm$ 0.004 & 0.084 & 856\\
$W_{gi}$ & -0.377 $\pm$ 0.008 & -2.62 $\pm$ 0.018 & -0.084 $\pm$ 0.004 & 0.088 & 856\\
$W_{iz}$ & 0.049 $\pm$ 0.012 & -2.223 $\pm$ 0.028 & 0.014 $\pm$ 0.007 & 0.135 & 856\\
$W_{gy}$ & -0.153 $\pm$ 0.01 & -2.399 $\pm$ 0.023 & -0.04 $\pm$ 0.006 & 0.113 & 856\\
\hline
\multicolumn{6}{c}{Convection set D}\\
\hline
$W_{ug}$ & -3.444 $\pm$ 0.021 & -2.798 $\pm$ 0.049 & -0.735 $\pm$ 0.013 & 0.233 & 711\\
$W_{ur}$ & -1.803 $\pm$ 0.012 & -2.773 $\pm$ 0.029 & -0.401 $\pm$ 0.008 & 0.138 & 711\\
$W_{gr}$ & -0.681 $\pm$ 0.008 & -2.753 $\pm$ 0.018 & -0.173 $\pm$ 0.005 & 0.088 & 711\\
$W_{gi}$ & -0.377 $\pm$ 0.008 & -2.571 $\pm$ 0.019 & -0.082 $\pm$ 0.005 & 0.09 & 711\\
$W_{iz}$ & 0.052 $\pm$ 0.012 & -2.177 $\pm$ 0.028 & 0.016 $\pm$ 0.007 & 0.131 & 711\\
$W_{gy}$ & -0.152 $\pm$ 0.01 & -2.35 $\pm$ 0.023 & -0.038 $\pm$ 0.006 & 0.111 & 711\\
\hline
\end{tabular}}
\label{tab:PWZ}
\end{table}

\begin{table}[ht!]
\caption{$PWZ$ relations for BL~Her models of the mathematical form $W_{\lambda_2, \lambda_1}=\alpha+\beta\log(P)+\gamma\mathrm{[Fe/H]}$ in low- and high-metallicity regimes for different wavelengths using different convective parameter sets.}
\centering
\scalebox{0.8}{
\begin{tabular}{c c c c c c}
\hline\hline
Band & $\alpha$ & $\beta$ & $\gamma$ & $\sigma$ & $N$\\
\hline \hline
\multicolumn{6}{c}{Low-metallicity regime ($Z=0.00135, 0.00061, 0.00043, 0.00014$)}\\
\hline
\multicolumn{6}{c}{Convection set A}\\
\hline
$W_{ug}$ & -3.286 $\pm$ 0.028 & -2.299 $\pm$ 0.041 & -0.325 $\pm$ 0.017 & 0.252 & 1734\\
$W_{ur}$ & -1.759 $\pm$ 0.019 & -2.569 $\pm$ 0.028 & -0.179 $\pm$ 0.012 & 0.173 & 1734\\
$W_{gr}$ & -0.715 $\pm$ 0.016 & -2.752 $\pm$ 0.023 & -0.079 $\pm$ 0.01 & 0.142 & 1734\\
$W_{gi}$ & -0.478 $\pm$ 0.016 & -2.653 $\pm$ 0.024 & -0.037 $\pm$ 0.01 & 0.148 & 1734\\
$W_{iz}$ & -0.153 $\pm$ 0.02 & -2.38 $\pm$ 0.029 & 0.008 $\pm$ 0.012 & 0.182 & 1734\\
$W_{gy}$ & -0.307 $\pm$ 0.018 & -2.496 $\pm$ 0.027 & -0.019 $\pm$ 0.011 & 0.166 & 1734\\
\hline
\multicolumn{6}{c}{Convection set B}\\
\hline
$W_{ug}$ & -3.293 $\pm$ 0.029 & -2.092 $\pm$ 0.041 & -0.335 $\pm$ 0.018 & 0.228 & 1217\\
$W_{ur}$ & -1.778 $\pm$ 0.021 & -2.41 $\pm$ 0.03 & -0.182 $\pm$ 0.013 & 0.166 & 1217\\
$W_{gr}$ & -0.742 $\pm$ 0.018 & -2.626 $\pm$ 0.025 & -0.078 $\pm$ 0.011 & 0.142 & 1217\\
$W_{gi}$ & -0.5 $\pm$ 0.019 & -2.517 $\pm$ 0.026 & -0.034 $\pm$ 0.012 & 0.146 & 1217\\
$W_{iz}$ & -0.159 $\pm$ 0.022 & -2.204 $\pm$ 0.03 & 0.013 $\pm$ 0.014 & 0.17 & 1217\\
$W_{gy}$ & -0.32 $\pm$ 0.021 & -2.336 $\pm$ 0.028 & -0.015 $\pm$ 0.013 & 0.159 & 1217\\
\hline
\multicolumn{6}{c}{Convection set C}\\
\hline
$W_{ug}$ & -3.258 $\pm$ 0.03 & -2.288 $\pm$ 0.041 & -0.372 $\pm$ 0.018 & 0.234 & 1292\\
$W_{ur}$ & -1.769 $\pm$ 0.022 & -2.553 $\pm$ 0.03 & -0.206 $\pm$ 0.013 & 0.169 & 1292\\
$W_{gr}$ & -0.752 $\pm$ 0.018 & -2.733 $\pm$ 0.025 & -0.092 $\pm$ 0.011 & 0.143 & 1292\\
$W_{gi}$ & -0.502 $\pm$ 0.019 & -2.601 $\pm$ 0.027 & -0.044 $\pm$ 0.012 & 0.15 & 1292\\
$W_{iz}$ & -0.137 $\pm$ 0.023 & -2.264 $\pm$ 0.032 & 0.008 $\pm$ 0.014 & 0.18 & 1292\\
$W_{gy}$ & -0.307 $\pm$ 0.021 & -2.413 $\pm$ 0.029 & -0.022 $\pm$ 0.013 & 0.165 & 1292\\
\hline
\multicolumn{6}{c}{Convection set D}\\
\hline
$W_{ug}$ & -3.276 $\pm$ 0.032 & -2.143 $\pm$ 0.041 & -0.378 $\pm$ 0.02 & 0.23 & 1094\\
$W_{ur}$ & -1.78 $\pm$ 0.024 & -2.432 $\pm$ 0.031 & -0.203 $\pm$ 0.015 & 0.173 & 1094\\
$W_{gr}$ & -0.758 $\pm$ 0.021 & -2.627 $\pm$ 0.027 & -0.084 $\pm$ 0.013 & 0.149 & 1094\\
$W_{gi}$ & -0.506 $\pm$ 0.021 & -2.502 $\pm$ 0.028 & -0.037 $\pm$ 0.013 & 0.154 & 1094\\
$W_{iz}$ & -0.139 $\pm$ 0.024 & -2.165 $\pm$ 0.032 & 0.013 $\pm$ 0.015 & 0.177 & 1094\\
$W_{gy}$ & -0.31 $\pm$ 0.023 & -2.312 $\pm$ 0.03 & -0.017 $\pm$ 0.014 & 0.166 & 1094\\
\hline
\multicolumn{6}{c}{High-metallicity regime ($Z=0.01300, 0.00834, 0.00424$)}\\
\hline
\multicolumn{6}{c}{Convection set A}\\
\hline
$W_{ug}$ & -3.472 $\pm$ 0.01 & -3.339 $\pm$ 0.025 & -0.973 $\pm$ 0.019 & 0.152 & 1532\\
$W_{ur}$ & -1.868 $\pm$ 0.009 & -3.12 $\pm$ 0.022 & -0.532 $\pm$ 0.017 & 0.137 & 1532\\
$W_{gr}$ & -0.771 $\pm$ 0.009 & -2.968 $\pm$ 0.022 & -0.23 $\pm$ 0.017 & 0.136 & 1532\\
$W_{gi}$ & -0.49 $\pm$ 0.01 & -2.783 $\pm$ 0.023 & -0.097 $\pm$ 0.018 & 0.142 & 1532\\
$W_{iz}$ & -0.107 $\pm$ 0.012 & -2.425 $\pm$ 0.029 & 0.057 $\pm$ 0.023 & 0.179 & 1532\\
$W_{gy}$ & -0.293 $\pm$ 0.011 & -2.583 $\pm$ 0.026 & -0.026 $\pm$ 0.02 & 0.159 & 1532\\
\hline
\multicolumn{6}{c}{Convection set B}\\
\hline
$W_{ug}$ & -3.504 $\pm$ 0.012 & -3.264 $\pm$ 0.031 & -1.05 $\pm$ 0.024 & 0.157 & 1043\\
$W_{ur}$ & -1.901 $\pm$ 0.011 & -3.044 $\pm$ 0.028 & -0.579 $\pm$ 0.021 & 0.14 & 1043\\
$W_{gr}$ & -0.806 $\pm$ 0.011 & -2.892 $\pm$ 0.027 & -0.257 $\pm$ 0.021 & 0.138 & 1043\\
$W_{gi}$ & -0.523 $\pm$ 0.011 & -2.68 $\pm$ 0.028 & -0.126 $\pm$ 0.022 & 0.144 & 1043\\
$W_{iz}$ & -0.138 $\pm$ 0.014 & -2.254 $\pm$ 0.035 & 0.015 $\pm$ 0.027 & 0.177 & 1043\\
$W_{gy}$ & -0.325 $\pm$ 0.012 & -2.441 $\pm$ 0.031 & -0.063 $\pm$ 0.024 & 0.16 & 1043\\
\hline
\multicolumn{6}{c}{Convection set C}\\
\hline
$W_{ug}$ & -3.444 $\pm$ 0.011 & -3.634 $\pm$ 0.025 & -1.132 $\pm$ 0.021 & 0.151 & 1340\\
$W_{ur}$ & -1.877 $\pm$ 0.01 & -3.246 $\pm$ 0.023 & -0.615 $\pm$ 0.019 & 0.136 & 1340\\
$W_{gr}$ & -0.805 $\pm$ 0.01 & -2.979 $\pm$ 0.023 & -0.262 $\pm$ 0.019 & 0.136 & 1340\\
$W_{gi}$ & -0.506 $\pm$ 0.01 & -2.768 $\pm$ 0.024 & -0.12 $\pm$ 0.02 & 0.145 & 1340\\
$W_{iz}$ & -0.07 $\pm$ 0.013 & -2.402 $\pm$ 0.031 & 0.035 $\pm$ 0.025 & 0.186 & 1340\\
$W_{gy}$ & -0.277 $\pm$ 0.012 & -2.566 $\pm$ 0.028 & -0.047 $\pm$ 0.023 & 0.166 & 1340\\
\hline
\multicolumn{6}{c}{Convection set D}\\
\hline
$W_{ug}$ & -3.448 $\pm$ 0.012 & -3.572 $\pm$ 0.03 & -1.169 $\pm$ 0.024 & 0.154 & 1028\\
$W_{ur}$ & -1.879 $\pm$ 0.011 & -3.198 $\pm$ 0.027 & -0.636 $\pm$ 0.022 & 0.139 & 1028\\
$W_{gr}$ & -0.807 $\pm$ 0.011 & -2.94 $\pm$ 0.027 & -0.272 $\pm$ 0.022 & 0.139 & 1028\\
$W_{gi}$ & -0.512 $\pm$ 0.012 & -2.709 $\pm$ 0.028 & -0.135 $\pm$ 0.023 & 0.147 & 1028\\
$W_{iz}$ & -0.09 $\pm$ 0.015 & -2.289 $\pm$ 0.036 & 0.005 $\pm$ 0.029 & 0.186 & 1028\\
$W_{gy}$ & -0.291 $\pm$ 0.013 & -2.476 $\pm$ 0.032 & -0.071 $\pm$ 0.026 & 0.166 & 1028\\
\hline
\end{tabular}}
\label{tab:PWZ_metallicityregime}
\end{table}

\subsection{Effect of convection parameters}

We use the standard $t$-test\footnote{We defined a $T$ statistic for the comparison of two linear regression slopes, $\hat{W}$ with sample sizes, $n$ and $m$, respectively \citep{ngeow2015}:
\begin{equation}\nonumber
T=\frac{\hat{W}_n-\hat{W}_m}{\sqrt{\mathrm{Var}(\hat{W}_n)+\mathrm{Var}(\hat{W}_m)}},
\label{eq:ttest}
\end{equation}
where $\mathrm{Var}(\hat{W})$ is the variance of the slope. The null hypothesis of equivalent slopes is rejected if $T>t_{\alpha/2,\nu}$ (or the probability of the observed value of the $T$ statistic) is $p<0.05,$ where $t_{\alpha/2,\nu}$ is the critical value under the two-tailed $t$-distribution with 95\% confidence limit ($\alpha$=0.05) and degrees of freedom, $\nu=n+m-4$.} to statistically compare the slopes of the theoretical $PL$ and $PW$ relations across the different convection parameter sets to study the effect of convection efficiencies on these relations. Note that the $t$-test is relatively robust to deviations from assumptions \citep[see e.g.][]{posten1984}.

The theoretical $PL$ and $PW$ relations for the BL~Her models in the Rubin--LSST filters and the results from the standard $t$-test across the different convection parameter sets are presented in Table~\ref{tab:PL} and Table~\ref{tab:PW}, respectively. For the complete set of BL~Her models, the slopes of the $PL$ and $PW$ relations for the models computed with radiative cooling (sets B and D) are statistically similar across all the Rubin--LSST filters longer than the $u$ filter. We found similar results in \citetalias{das2021} for the Johnson-Cousin-Glass bands ($BVRIJHKLL'M$) and in \citetalias{das2024} across all the $Gaia$ passbands. However, for the low-mass BL~Her models, the theoretical $PL$ relations computed using convection parameter sets A, C, and D exhibit statistically similar slopes across most of the Rubin--LSST filters while the $PW$ slopes are only statistically similar for models computed with sets A and D. This result is similar to what we found in \citetalias{das2024} for the $Gaia$ passbands.

In increasing order of their central effective wavelengths ($\lambda_{\rm eff}$), the Rubin--LSST filters follow the order of $u<g<r<i<z<y$ \citep{rodrigo2012, rodrigo2020}. Table~\ref{tab:PL} demonstrates that theoretical $PL$ slopes get steeper in the Rubin--LSST filters with increasing wavelengths, similar to what was found in \citetalias{das2021} for the Johnson-Cousin-Glass bands ($UBVRIJHKLL'M$) and in \citetalias{das2024} for the $Gaia$ passbands ($GG_{BP}G_{RP}$). This trend is also observed in empirical $PL$ relations for RR~Lyrae stars \citep[][among others]{neeley2017, beaton2018b, bhardwaj2020a}. The dispersion in the theoretical $PL$ relations for BL~Her models also decreases with increasing wavelengths in the Rubin--LSST filters and is expected to be caused by the very decrease in the width of the instability strip itself with increasing wavelengths \citep[for more details, see,][]{catelan2004, madore2012, marconi2015}.

\begin{figure*}
\centering
\includegraphics[scale = 1]{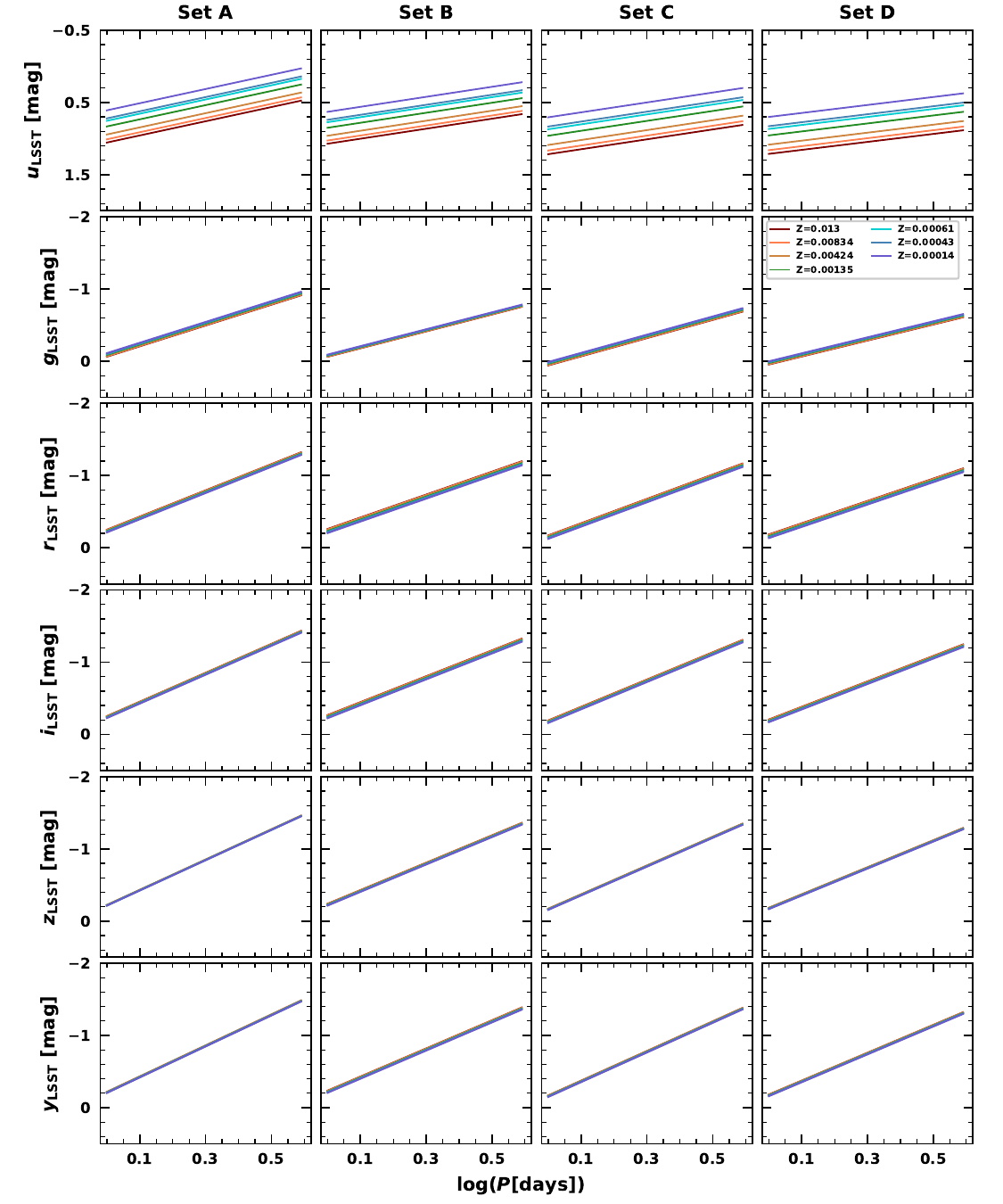}
\caption{The predicted multi-filter $PL$ relations of the BL~Her models with different chemical compositions across the different Rubin-LSST wavelengths for the convective parameter sets~A, B, C, and D.  The $y$-scale is same (2.5 mag) in each panel for a relative comparison.}
\label{fig:PLZ}
\end{figure*}

\begin{figure*}
\centering
\includegraphics[scale = 1]{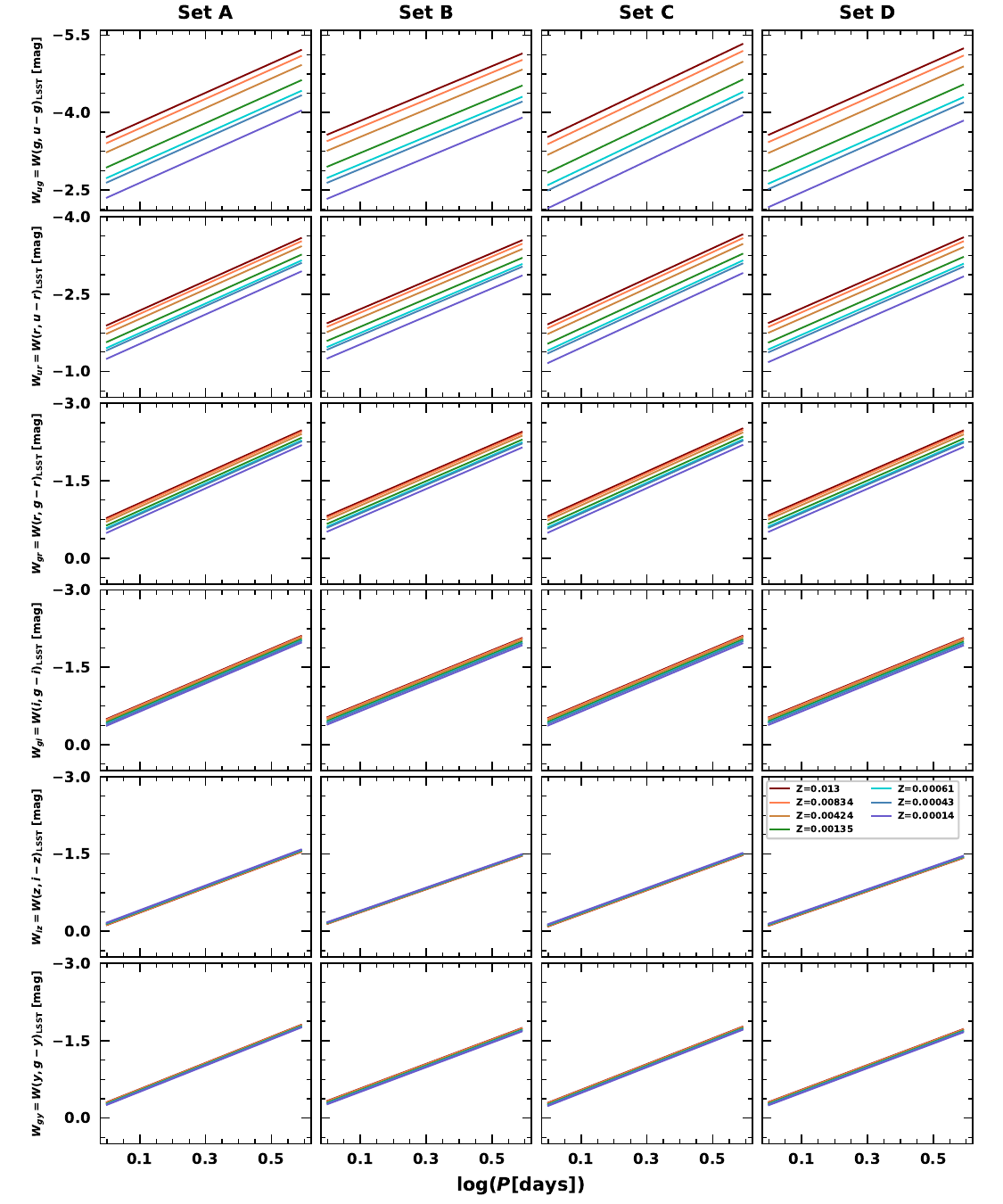}
\caption{The predicted multi-filter period-Wesenheit ($W_{ug}, W_{ur}, W_{gr}, W_{gi},  W_{iz}, W_{gy}$) relations of the BL~Her models with different chemical compositions across the different Rubin-LSST wavelengths for the convective parameter sets~A, B, C, and D.  The $y$-scale is same (3.5 mag) in each panel for a relative comparison.}
\label{fig:PWZ}
\end{figure*}

\begin{figure*}
\centering
\includegraphics[scale = 0.95]{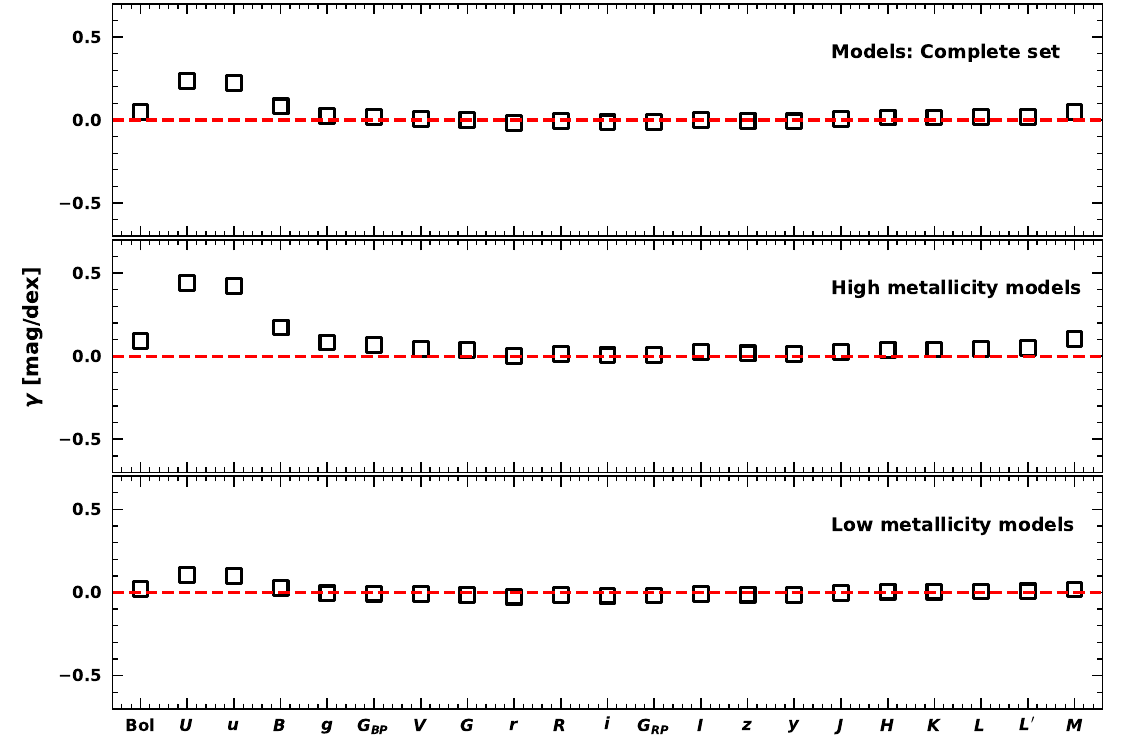}
\caption{Contribution of the $\gamma$ term (metallicity effect) obtained from the theoretical $PLZ$ relations for BL~Her models computed using the convection parameter set~A. Figure updated from \citetalias{das2024} to also include the Rubin--LSST filters ($ugrizy$) in addition to the bolometric (Bol), Johnson-Cousins-Glass ($UBVRIJHKLL'M$) and $Gaia$ ($GG_{BP}G_{RP}$) passbands. The different panels present results from the complete set of models compared with the high-metallicity models ($Z=0.00424, 0.00834, 0.01300$) and the low-metallicity models ($Z=0.00014, 0.00043, 0.00061,  0.00135$). The x-axis is in increasing order of the central effective wavelengths ($\lambda_{\rm eff}$) of the respective passbands as provided by the SVO Filter Profile Service \citep{rodrigo2012, rodrigo2020}.}
\label{fig:gamma}
\end{figure*}

\subsection{Effect of metallicity}

As mentioned earlier, the weak or negligible metallicity effect on the $PL$ relations of BL~Her stars from theoretical and empirical studies so far seem to offer an advantage over the classical pulsators. RR~Lyrae and classical Cepheid stars suffer from strong metallicity effects on their $PL$ relations, especially at shorter wavelengths \citep[see, for example,][]{ripepi2021, ripepi2022, breuval2022, desomma2022, bhardwaj2023}. BL~Her stars thereby potentially exhibit a universal $PL$ relation irrespective of different stellar environments with different metallicities. It is therefore of crucial importance to also test the effect of metallicity on the theoretical relations in the Rubin--LSST filters. To this end, we derive $PLZ$ and $PWZ$ relations for the BL~Her models in the Rubin--LSST filters of the mathematical form:
\begin{equation}
\begin{aligned}
M_\lambda =& \alpha+\beta\log(P)+\gamma\mathrm{[Fe/H]}; \\
W_{\lambda_2, \lambda_1} =& \alpha+\beta\log(P)+\gamma\mathrm{[Fe/H]}.\\
\end{aligned}
\label{eq:PLZ}
\end{equation}

The theoretical $PLZ$ relations of the BL~Her models for the different Rubin--LSST filters across the different convection parameter sets are presented in Tables~\ref{tab:PLZ} and \ref{tab:PLZ_metallicityregime} and displayed in Fig.~\ref{fig:PLZ}. In Table~\ref{tab:PLZ}, the $PLZ$ relations from both the complete set of models and from the low-mass models exhibit weak or negligible effect of metallicity (indicated by the $\gamma$ term) on the $PL$ relations for wavelengths longer than the $g$ filter onwards (within 3$\sigma$ uncertainties), with significant metallicity effect in the $u$ filter. This is very similar to the strong dependence of $PL$ relations on metallicity that was observed in the $U$ and $B$ bands in \citetalias{das2021}. To investigate the significant metallicity dependence in the $u$ filter, we derived $PLZ$ relations for the BL~Her models in Table~\ref{tab:PLZ_metallicityregime} in the low-metallicity ($Z=0.00014, 0.00043, 0.00061, 0.00135$) and the high-metallicity ($Z=0.00424, 0.00834, 0.01300$) regimes. We clearly see the significant metallicity contribution in the $u$ filter arising mostly from the high-metallicity BL~Her models with a much smaller dependence on metallicity from the low-metallicity BL~Her models. This could be caused by the higher effect of metallicities on bolometric corrections at wavelengths shorter than the $V$ band \citep{gray2005, kudritzki2008} and is discussed in more details in \citetalias{das2021} and \citetalias{das2024}. A systematic study of the impact of the different model atmospheres adopted for the transformations of the bolometric light curves is crucial for constraining the stellar pulsation theory further, but is beyond the scope of the present study.

The equivalent multi-filter $PWZ$ relations for the BL~Her models are derived in Tables~\ref{tab:PWZ} and \ref{tab:PWZ_metallicityregime} and displayed in Fig.~\ref{fig:PWZ}. Strong metallicity effects are observed in the $PWZ$ relations involving the $u$ filter and are once again, significantly contributed by the high-metallicity BL~Her models. The small but significant metallicity contribution in the $W_{gr}$ relations when there exists weak or negligible metallicity effect on the $PL$ relations in the $g$ and $r$ filters individually is interesting but is similar to what was also seen in \citetalias{das2024} for $PW$ relations derived using $Gaia$ passbands. The $PW_{Gaia}Z$ relations from the classical Cepheids also exhibit a relatively large metallicity dependence \citep[see][]{ripepi2022, breuval2022, trentin2024}. As in \citetalias{das2024}, the high-metallicity BL~Her models exhibit a small but significant and negative metallicity coefficient in the $PW_{gr}Z$ relations, resulting in an overall small metallicity effect on the $PW_{gr}Z$ relations. An updated figure demonstrating the contribution of the $\gamma$ term (metallicity effect) from the the theoretical $PLZ$ relations for BL~Her models across multiple wavelengths, including the Rubin--LSST filters ($ugrizy$) is presented in Fig.\ref{fig:gamma}. This figure clearly exhibits weak or negligible metallicity effect on the $PL$ relations for BL~Her models for wavelengths longer than $V$ band. In addition, for wavelengths shorter than $V$ band, we find a much higher contribution of metallicity from the high-metallicity BL~Her models that contribute to the overall metallicity dependence when the complete set of BL~Her models is considered.

In addition to Eq.~\ref{eq:PLZ}, we also derive \textit{modified} $PLZ$ and $PWZ$ relations for the BL~Her models in the Rubin--LSST filters of the mathematical form:
\begin{equation}
\begin{aligned}
M_\lambda =& \alpha+(\beta_1+\beta_2[Fe/H])\log(P)+\gamma\mathrm{[Fe/H]}; \\
W_{\lambda_2, \lambda_1} =& \alpha+(\beta_1+\beta_2[Fe/H])\log(P)+\gamma\mathrm{[Fe/H]}\\
\end{aligned}
\end{equation}
to take into account metallicity-dependent slopes that is predicted from the theoretical $PL$ and $PW$ relations involving the Rubin--LSST $u$ filter. The modified $PLZ$ and $PWZ$ relations are presented in Tables~\ref{tab:PLZ_modified} and \ref{tab:PWZ_modified}, respectively in the Appendix. For completeness, we also include the relations for all the Rubin--LSST filters. The dependence on metallicity for the Rubin--LSST $u$ filter indeed seems to be weaker (but still significant considering 3$\sigma$ uncertainties) when we use the modified $PLZ/PWZ$ relations. Comparing results from the complete set of models using convection set~A, we find that the metallicity term $\gamma$ from the $PLZ$ relation is $0.223 \pm 0.009$ (Table~\ref{tab:PLZ}) while from the \textit{modified} $PLZ$ relation, it is $0.125 \pm 0.02$ (Table~\ref{tab:PLZ_modified}) in the Rubin--LSST $u$ filter. Similar results are seen for the other convection sets as well as for the case of low-mass models only. This is also true for the $PWZ$ relations involving the Rubin--LSST $u$ filter for all the cases.

As was observed for the theoretical $PW$ relations for RR~Lyrae models in \citet{marconi2022}, a dependence of the metallicity effect on the band combinations is seen for BL~Her models in this work. The metallicity effect is minimal for relations involving the Wesenheit indices $W_{gi}$, $W_{iz}$ and $W_{gy}$ and we therefore recommend using these filter combinations for BL~Her stars to be used as reliable standard candles, when observed with the Rubin--LSST. This will take care of any uncertainties present in the spectroscopic metallicities of these stars.

\section{Comparison with RRL models}
\label{sec:RRL}

Recent empirical studies by \citet{majaess2010}, \citet{bhardwaj2017b} and \citet{braga2020} have found evidence that RRab stars and T2Cs obey the same $PL$ relations at near-infrared wavelengths. This is not surprising given that both classes are population~II classical pulsators having similar chemical compositions, with the classical evolutionary scenario often suggesting BL~Her stars to be the evolved counterparts of RR~Lyrae stars as they evolve from the blue Zero Age Horizontal Branch and approach their AGB tracks \citep{gingold1985, marconi2011, marconi2015, bono2020}. In \citetalias{das2021}, we had found $PL$ slopes from our grid of BL~Her models computed without radiative cooling (sets A and C) to be statistically similar with those for RR~Lyrae models from \citet{marconi2015} in the $RIJHK_S$ bands. In addition, we had also found the period-radius slopes of our grid of BL~Her models to be similar with those for RR~Lyrae models from \citet{marconi2015} across all four sets of convection parameters, thereby suggesting a tight coupling of both pulsational and evolutionary properties between the two classes of low-mass radial pulsators. It is important to note that this similarity between RR~Lyraes and BL~Her stars may not hold true over the entire metallicity range as cautioned by \citet{marconi2015}. However, RR~Lyraes and T2Cs potentially following the same $PL$ relations offers us a unique opportunity of adopting RRLs+T2Cs together for the calibration of the extragalactic distance scale (and thereby estimating the local Hubble constant) as an alternative to classical Cepheids \citep[see, e.g.][]{beaton2016}.

First, we compare the stellar parameter space of our grid 
to that of the RR~Lyrae models from \citet{marconi2015} which have been subsequently used in \citet{marconi2022} and derive $MLTZP$ relations similar to Eq.~1 of \citet{marconi2015}; the relations are presented in Table~\ref{tab:MLTZP}. We highlight here that while \citet{marconi2022} uses pulsation models constrained by stellar evolution tracks, our grid of BL~Her pulsation models have so far not been constrained by stellar evolutionary tracks, which may influence this comparison.

\begin{table*}
\caption{Comparison of the pulsation period and its dependence on stellar luminosity, stellar mass, effective temperature and chemical composition of the mathematical form $\log(P)=a+b\log(L/L_{\odot})+c\log(M/M_{\odot})+d\log(T_{\rm eff})+e\log(Z)$ for our grid of BL~Her models computed using four sets of convection parameters and for fundamental-mode RR~Lyrae models from \citet{marconi2015}.}
\centering
\scalebox{1}{
\begin{tabular}{c c c c c c}
\hline
Source & a & b & c & d & e \\
\hline
BL~Her (Set~A) & 13.458 $\pm$ 0.009 & 0.869 $\pm$ 0.000 & -0.731 $\pm$ 0.001 &  -4.004 $\pm$ 0.002 & 0.009 $\pm$ 0.000 \\
BL~Her (Set~B) & 13.587 $\pm$ 0.015 & 0.874 $\pm$ 0.001 & -0.746 $\pm$ 0.002 & -4.042 $\pm$  0.004 & 0.008 $\pm$ 0.000\\
BL~Her (Set~C) & 14.445 $\pm$ 0.025 & 0.865 $\pm$ 0.001 & -0.745 $\pm$ 0.003 & -4.265 $\pm$  0.006 & 0.006 $\pm$ 0.000\\
BL~Her (Set~D) &14.616 $\pm$ 0.054 & 0.857 $\pm$ 0.002 & -0.739 $\pm$ 0.005 & -4.305 $\pm$  0.014 & 0.007 $\pm$ 0.000\\
RRab & 11.347 $\pm$ 0.006 & 0.860 $\pm$ 0.003 & -0.580 $\pm$ 0.020 &  -3.430 $\pm$ 0.010 & 0.024 $\pm$ 0.002\\
\hline
\end{tabular}}
\label{tab:MLTZP}
\end{table*}

\begin{figure*}
\centering
\includegraphics[scale = 1]{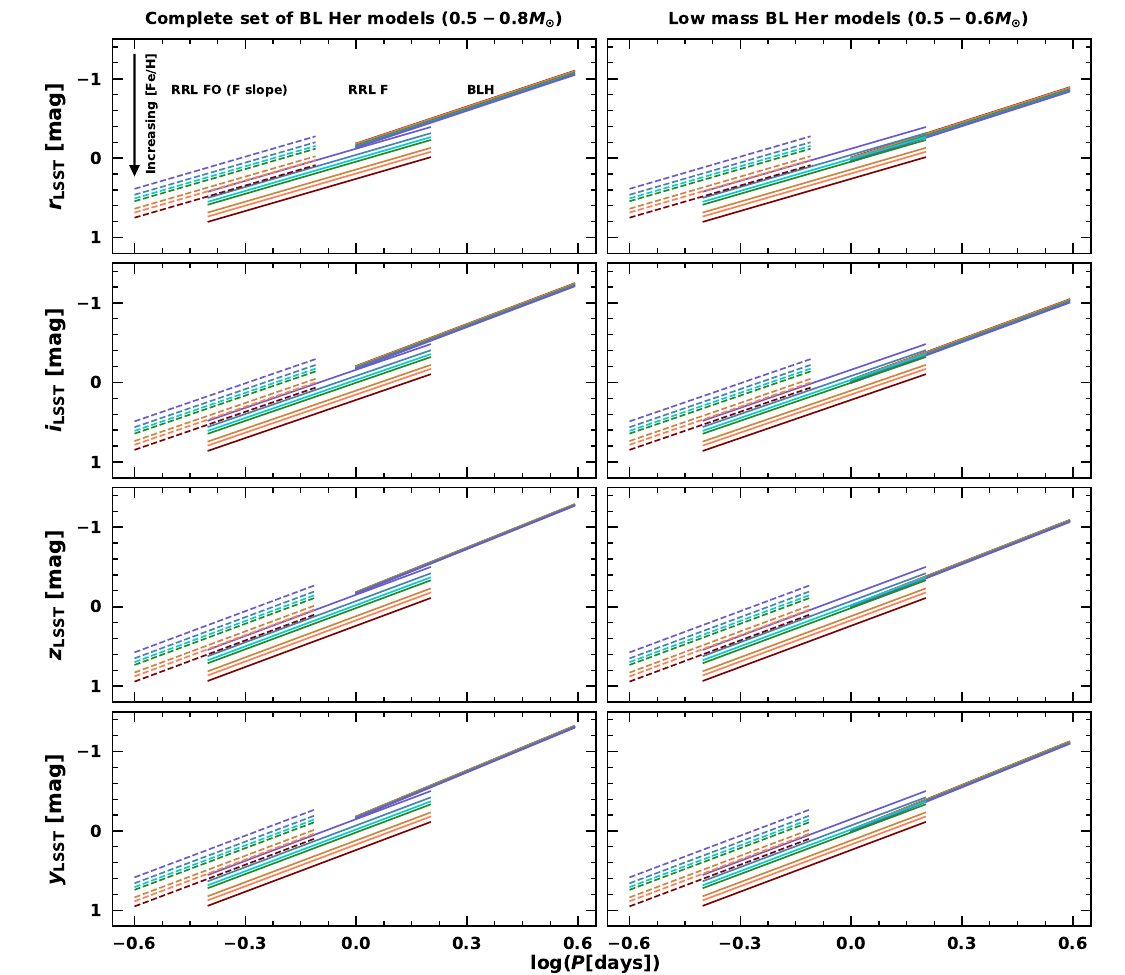}
\caption{A comparison of the theoretical $PL$ slopes for RR~Lyrae models from \citet{marconi2022} and for BL~Her models using convection parameter set~D from this work in the Rubin--LSST $rizy$ passbands.}
\label{fig:PLZ_RRLBLH}
\end{figure*}

We thereby compare the theoretical $PL$ and $PW$ relations of the RR~Lyrae models from \citet{marconi2022} with those from our BL~Her models in the Rubin--LSST filters. As summarized in Table~\ref{tab:RRL_PL}, we find the theoretical $PL$ slopes of the RR~Lyrae models from \citet{marconi2022} to be statistically similar to those from our low-mass BL~Her models across the four different convection parameter sets in the Rubin--LSST $rizy$ filters. This is also displayed in Fig.~\ref{fig:PLZ_RRLBLH} for the BL~Her models computed using convection parameter set~D; similar results hold for the other convection sets. However, this result does not hold true when we consider the complete set of BL~Her models with stellar masses $M/M_{\odot} \in [0.5, 0.8]$. Note that this could be because the complete set of BL~Her models also includes stellar masses higher than what is typically considered for BL~Her stars from stellar evolutionary predictions, as discussed in Section~\ref{sec:data}. A similar comparison for the $PWZ$ relations is presented in Table~\ref{tab:RRL_PW}. While the $PWZ$ slopes are still statistically different when comparing the complete set of BL~Her models to the RR~Lyrae models, there are a few cases where the slopes are predicted to be statistically similar when considering the low-mass BL~Her models only. Note that there are several reasons why the theoretical relations for RR~Lyrae models may differ from the corresponding relations for BL~Her models from this work. First, the two studies use different theories of convection-- \citet{marconi2022} uses the convection formulation outlined in \citet{stellingwerf1982a, stellingwerf1982b} while this work uses \textsc{mesa-rsp} which follows the \citet{kuhfuss1986} turbulent convection theory. Next, there is further uncertainty involved with the different model atmospheres (and thus, bolometric correction tables) used between the two studies. Lastly, note that this study uses the two-tailed $t$-distribution with quite a stringent confidence limit of 95\% for the statistical comparison of the slopes. 

\begin{table*}
\caption{Comparison of the slopes ($\beta$ = coefficient of the $\log(P)$ term) of the $PLZ$ relations of the mathematical form $M_\lambda=\alpha+\beta\log(P)+\gamma\mathrm{[Fe/H]}$ for RR~Lyrae models from \citet{marconi2022} with those for BL~Her models computed in this work using the Rubin-LSST filters. $N$ is the total number of fundamental mode RR~Lyrae models. |$T$| represents the observed value of the $t$-statistic, and $p(t)$ gives the probability of acceptance of the null hypothesis (equal slopes). The bold-faced entries indicate that the null hypothesis of the equivalent slopes can be rejected.}
\centering
\scalebox{0.9}{
\begin{tabular}{c c c c c c c c c c c}
\hline
Band & Mode & $\alpha$ & $\beta$ & $\gamma$ & $\sigma$ & $N$& \multicolumn{4}{c}{(|$T$|, $p(t)$) w.r.t. BL~Her models from this work$^\dagger$}\\
(LSST) & & & & & & & Set A & Set B & Set C & Set D\\
\hline
\multicolumn{11}{c}{Complete set of models ($0.5-0.8M_{\odot}$)}\\
\hline
$r$ & F &0.25$\pm$0.02 & -1.35$\pm$0.08 &0.163$\pm$0.014 & 0.128& 155 & \textbf{(5.476,0.0)} & \textbf{(2.786,0.005)} & \textbf{(3.888,0.0)} & \textbf{(2.298,0.022)}\\
$i$ & F &0.21$\pm$0.02 & -1.6$\pm$0.07 &0.163$\pm$0.011 &0.099 &155 & \textbf{(5.357,0.0)} & \textbf{(2.586,0.01)} & \textbf{(3.807,0.0)} & \textbf{(2.112,0.035)}\\
$z$ & F &0.23$\pm$0.02& -1.73$\pm$0.06 &0.168$\pm$0.01 &0.087 &155 & \textbf{(5.648,0.0)} & \textbf{(2.554,0.011)} & \textbf{(4.012,0.0)} & \textbf{(2.114,0.035)}\\
$y$ & F &0.23$\pm$0.02 & -1.75$\pm$0.06 &0.167$\pm$0.01 &0.086 &155 & \textbf{(6.112,0.0)} & \textbf{(3.059,0.002)} & \textbf{(4.618,0.0)} & \textbf{(2.721,0.007)}\\
\hline
\multicolumn{11}{c}{Low-mass models only ($0.5-0.6M_{\odot}$)}\\
\hline
$r$ & F &0.25$\pm$0.02 & -1.35$\pm$0.08 &0.163$\pm$0.014 & 0.128& 155& (0.606,0.545) & (0.666,0.505) & (1.843,0.066) & (1.558,0.12)\\
$i$ & F &0.21$\pm$0.02 & -1.6$\pm$0.07 &0.163$\pm$0.011 &0.099 &155& (0.518,0.605) & (0.853,0.394) & (1.761,0.078) & (1.394,0.164)\\
$z$ & F &0.23$\pm$0.02& -1.73$\pm$0.06 &0.168$\pm$0.01 &0.087 &155& (0.389,0.698) & (1.155,0.248) & (1.787,0.074) & (1.315,0.189)\\
$y$ & F &0.23$\pm$0.02 & -1.75$\pm$0.06 &0.167$\pm$0.01 &0.086 &155& (0.964,0.335) & (0.581,0.561) & \textbf{(2.401,0.017)} & (1.872,0.062)\\
\hline
\end{tabular}}
\tablefoot{\small 
        $^\dagger$ The $PLZ$ relations for the BL~Her models are provided in Table~\ref{tab:PLZ}.}
\label{tab:RRL_PL}
\end{table*}

\begin{table*}
\caption{Same as Table~\ref{tab:RRL_PL} but for the $PWZ$ relations of the mathematical form $W_{\lambda_2, \lambda_1}=\alpha+\beta\log(P)+\gamma\mathrm{[Fe/H]}$ for RR~Lyrae models from \citet{marconi2022} with those for BL~Her models computed in this work using the Rubin-LSST filters.}
\centering
\scalebox{0.9}{
\begin{tabular}{c c c c c c c c c c c}
\hline
Band & Mode & $\alpha$ & $\beta$ & $\gamma$ & $\sigma$ & $N$& \multicolumn{4}{c}{(|$T$|, $p(t)$) w.r.t. BL~Her models from this work$^\dagger$}\\
(LSST) & & & & & & & Set A & Set B & Set C & Set D\\
\hline
\multicolumn{11}{c}{Complete set of models ($0.5-0.8M_{\odot}$)}\\
\hline
$W_{ug}$ & F &-3.28$\pm$0.04 & -1.29$\pm$0.14 &-0.22$\pm$0.02 &0.219 & 155 & \textbf{(10.968,0.0)}& \textbf{(9.491,0.0)} & \textbf{(12.25,0.0)} & \textbf{(10.7,0.0)}\\
$W_{ur}$ & F &-1.622$\pm$0.019 & -2.09$\pm$0.07 &-0.059$\pm$0.011 &0.010 & 155 & \textbf{(10.671,0.0)} & \textbf{(8.469,0.0)} & \textbf{(11.672,0.0)} & \textbf{(9.595,0.0)}\\
$W_{gr}$ & F &-0.489$\pm$0.013 & -2.63$\pm$0.04 &0.047$\pm$0.008 &0.068 & 155 & \textbf{(5.548,0.0)} & \textbf{(2.8,0.005)} & \textbf{(5.591,0.0)} & \textbf{(3.309,0.001)}\\
$W_{gi}$ & F &-0.178$\pm$0.009 & -2.51$\pm$0.03 &0.11$\pm$0.005 &0.046 & 155 & \textbf{(6.148,0.0)} & \textbf{(2.394,0.017)} & \textbf{(5.259,0.0)} & \textbf{(2.496,0.013)}\\
$W_{iz}$ & F &0.295$\pm$0.01 & -2.13$\pm$0.03 &0.185$\pm$0.006 &0.051 & 155 & \textbf{(7.373,0.0)} & \textbf{(2.566,0.01)} & \textbf{(5.537,0.0)} & \textbf{(2.395,0.017)}\\
$W_{gy}$ & F &0.071$\pm$0.008 & -2.23$\pm$0.03 &0.147$\pm$0.005 &0.042 & 155 & \textbf{(8.73,0.0)} & \textbf{(4.26,0.0)} & \textbf{(7.35,0.0)} & \textbf{(4.274,0.0)}\\
\hline
\multicolumn{11}{c}{Low-mass models only ($0.5-0.6M_{\odot}$)}\\
\hline
$W_{ug}$ & F &-3.28$\pm$0.04 & -1.29$\pm$0.14 &-0.22$\pm$0.02 &0.219 & 155& \textbf{(8.7,0.0)} & \textbf{(7.662,0.0)} & \textbf{(11.106,0.0)} & \textbf{(10.167,0.0)}\\
$W_{ur}$ & F &-1.622$\pm$0.019 & -2.09$\pm$0.07 &-0.059$\pm$0.011 &0.010 & 155& \textbf{(7.811,0.0)} & \textbf{(6.282,0.0)} & \textbf{(10.228,0.0)} & \textbf{(9.014,0.0)}\\
$W_{gr}$ & F &-0.489$\pm$0.013 & -2.63$\pm$0.04 &0.047$\pm$0.008 &0.068 & 155& \textbf{(2.646,0.008)} & (0.677,0.498) & \textbf{(4.188,0.0)} & \textbf{(2.804,0.005)}\\
$W_{gi}$ & F &-0.178$\pm$0.009 & -2.51$\pm$0.03 &0.11$\pm$0.005 &0.046 & 155& (1.479,0.139) & (0.972,0.331) & \textbf{(3.144,0.002)} & (1.718,0.086)\\
$W_{iz}$ & F &0.295$\pm$0.01 & -2.13$\pm$0.03 &0.185$\pm$0.006 &0.051 & 155& (0.0,1.0) & \textbf{(2.561,0.011)} & \textbf{(2.266,0.024)} & (1.145,0.252)\\
$W_{gy}$ & F &0.071$\pm$0.008 & -2.23$\pm$0.03 &0.147$\pm$0.005 &0.042 & 155& \textbf{(2.258,0.024)} & (0.269,0.788) & \textbf{(4.471,0.0)} & \textbf{(3.174,0.002)}\\
\hline
\end{tabular}}
\tablefoot{\small 
        $^\dagger$ The $PWZ$ relations for the BL~Her models are provided in Table~\ref{tab:PWZ}.}
\label{tab:RRL_PW}
\end{table*}

\section{Summary and Conclusion}
\label{sec:results}

The upcoming Rubin--LSST is expected to revolutionise the field of classical pulsators with unprecedented photometric data of the deep universe. We extended the grid of BL~Her models from \citetalias{das2021} and \citetalias{das2024}, computed using \textsc{mesa-rsp} \citep{smolec2008, paxton2019}, to obtain new theoretical light curves in the Rubin--LSST $ugrizy$ filters. The input stellar parameters ($ZXMLT_{\rm eff}$) were chosen typical for BL~Her stars and ranges over the values: metallicity ($-2.0\; \mathrm{dex} \leq \mathrm{[Fe/H]} \leq 0.0\; \mathrm{dex}$), stellar mass (0.5--0.8\,$M\,_{\odot}$), stellar luminosity (50--300\,$L_{\odot}$), and effective temperature (full extent of the instability strip in steps of 50\,K; 5900--7200\,K for 50\,$L_{\odot}$ and 4700--6550\,K for 300\,$L_{\odot}$). The grid of models was also computed for the four different sets of convection parameters as provided in \citet{paxton2019}. The BL~Her models used in this analysis were checked for attaining full-amplitude stable pulsations and for fundamental-mode pulsation with pulsation periods between 1 and 4 days, following the conventional period classification of \citet{soszynski2018}.

In this work, we derived new theoretical $PL$ and $PW$ relations for the fine grid of BL~Her models in the Rubin--LSST $ugrizy$ filters and thereby studied the effects of convection parameters and metallicity on these relations. We found the $PL$ and $PW$ slopes for the models computed with radiative cooling (sets B and D) to be statistically similar across the Rubin--LSST $grizy$ filters, similar to what was observed in \citetalias{das2021} for the Johnson-Cousin-Glass bands ($BVRIJHKLL'M$) and in \citetalias{das2024} for the $Gaia$ passbands ($GG_{BP}G_{RP}$). Considering the low-mass models only, the convection parameter sets A, C, and D exhibit statistically similar $PL$ slopes across most of the Rubin--LSST filters while only convection parameter sets A and D have statistically similar $PW$ slopes. The $PL$ relations of the BL~Her models exhibit steeper slopes but smaller dispersion with increasing wavelengths in the Rubin--LSST filters. 

It is important to note here that when comparing with future observations from the Rubin--LSST, in addition to comparing the $PL$ and $PW$ relations at mean light, one must also compare the light curve structures in the respective filters over a complete pulsation cycle to be able to distinguish among the preferred convection parameter sets better. This was recently demonstrated for the comparison of theoretical and empirical relations for BL~Her stars in the $Gaia$ passbands, where although the $PL$ and $PW$ relations at mean light did not prefer any convection parameter set \citepalias{das2024}, a clear preference for non-linear BL Her models computed without radiative cooling was observed when we compared model light curves to observations (for further discussion on this, see \citetalias{das2025}). Note that a known (but unsolved) potential shortcoming on the theoretical front is the need to adopt static model atmospheres even though the pulsating atmosphere is clearly dynamic. This may affect detailed comparisons between observed and theoretical light curves.

The BL~Her models exhibit weak or negligible effect of metallicity on the $PL$ relations for wavelengths longer than the $g$ filter for both the cases of the complete set of models as well as the low-mass models. However, we find significant metallicity effect on the $PL$ relation in the $u$ filter, similar to what was demonstrated for wavelengths shorter than $V$ band in \citetalias{das2021}.  Strong metallicity effects are observed in the $PWZ$ relations involving the $u$ filter and are found to have significant contribution from the high-metallicity BL~Her models. A possible explanation for the stronger contribution of metallicity to both the $PL$ and $PW$ relations from the high-metallicity BL~Her models could be the increased sensitivity of bolometric corrections to metallicities for filters with wavelengths shorter than the $V$ band \citep{gray2005, kudritzki2008}. The theoretical relations therefore indeed predict that BL~Her stars will obey almost metallicity-independent $PL$ relations at the longer wavelengths of the Rubin--LSST ($rizy$).

Lastly, following recent empirical studies \citep{majaess2010, bhardwaj2017b, braga2020} that suggest that RRab stars and T2Cs obey the same $PL$ relations at near-infrared wavelengths, we compare the multi-filter Rubin--LSST $PLZ$ and $PWZ$ relations for our BL~Her models with those for RR~Lyrae stars by \citet{marconi2022}. We found the theoretical $PL$ slopes of the RR~Lyrae models to be statistically similar with those from our low-mass BL~Her models across the four different convection parameter sets in the Rubin--LSST $rizy$ filters. Both the $PLZ$ and $PWZ$ relations exhibit statistically different slopes between the RR~Lyrae models and the BL~Her models when the complete set is considered with stellar masses $M/M_{\odot} \in [0.5, 0.8]$. Note that this could be because the complete set of BL~Her models also includes stellar masses higher than what is typically considered for BL~Her stars from stellar evolutionary predictions. We highlight here that the two studies use different theories of convection as well as different assumptions of model atmospheres which could contribute to different slopes for a few certain cases when comparing the $PWZ$ relations of the low-mass BL~Her models with the RR~Lyrae models.

While our fine grid of BL~Her models ushers in the era of large number statistics in stellar pulsation modelling, we point out that our grid uses the four convection parameter sets as outlined in \citet{paxton2019} which are useful starting choices but may need further calibration to match with observed light curves in the Rubin--LSST filters over an entire pulsation cycle. In addition, a systematic and robust study of the impact of different model atmospheres adopted for the transformations of the bolometric light curves will be crucial for constraining the stellar pulsation theory further.

\begin{acknowledgements}
The authors thank the referee for useful comments and suggestions that improved the quality of the manuscript. This research was supported by the KKP-137523 `SeismoLab' \'Elvonal grant, the SNN-147362 grant, and the NKFIH excellence grant TKP2021-NKTA-64 of the Hungarian Research, Development and Innovation Office (NKFIH). RS is supported by the Polish National Science Centre, SONATA BIS grant, 2018/30/E/ST9/00598. This research was supported by the International Space Science Institute (ISSI) in Bern/Beijing through ISSI/ISSI-BJ International Team project ID \#24-603 - “EXPANDING Universe” (EXploiting Precision AstroNomical Distance INdicators in the Gaia Universe). The authors acknowledge the use of High Performance Computing facility Pegasus at IUCAA, Pune as well as at the HUN-REN (formerly ELKH) Cloud and the following software used in this project: \textsc{mesa}~r11701 \citep{paxton2011, paxton2013, paxton2015, paxton2018, paxton2019, jermyn2023}. This research has made use of NASA’s Astrophysics Data System.
\end{acknowledgements}

\bibliographystyle{aa} 

\begin{appendix}

\section{Chemical compositions of the BL~Her models}
In Table~\ref{tab:composition}, we present the equivalent $ZX$ values for each [Fe/H] value.
\begin{table}
\caption{Chemical compositions of the adopted pulsation models.}
\centering
\begin{tabular}{c c c}
\hline
[Fe/H] & $Z$ & $X$\\
\hline
-2.00 & 0.00014 & 0.75115\\
-1.50 & 0.00043 & 0.75041\\
-1.35 &0.00061& 0.74996\\
-1.00 &0.00135& 0.74806\\
-0.50&0.00424&0.74073\\
-0.20 &0.00834& 0.73032\\
0.00 &0.01300& 0.71847\\
\hline
\end{tabular}
\tablefoot{\small
	 The $Z$ and $X$ values are estimated from the $\mathrm{[Fe/H]}$ values by assuming the primordial helium value of 0.2485 from the WMAP CMB observations \citep{hinshaw2013} and the helium enrichment parameter value of 1.54 \citep{asplund2009}. The solar mixture is adopted from \citet{asplund2009}.}
\label{tab:composition}
\end{table}

\section{Bailey diagram}
\begin{figure*}
\centering
\includegraphics[scale = 0.95]{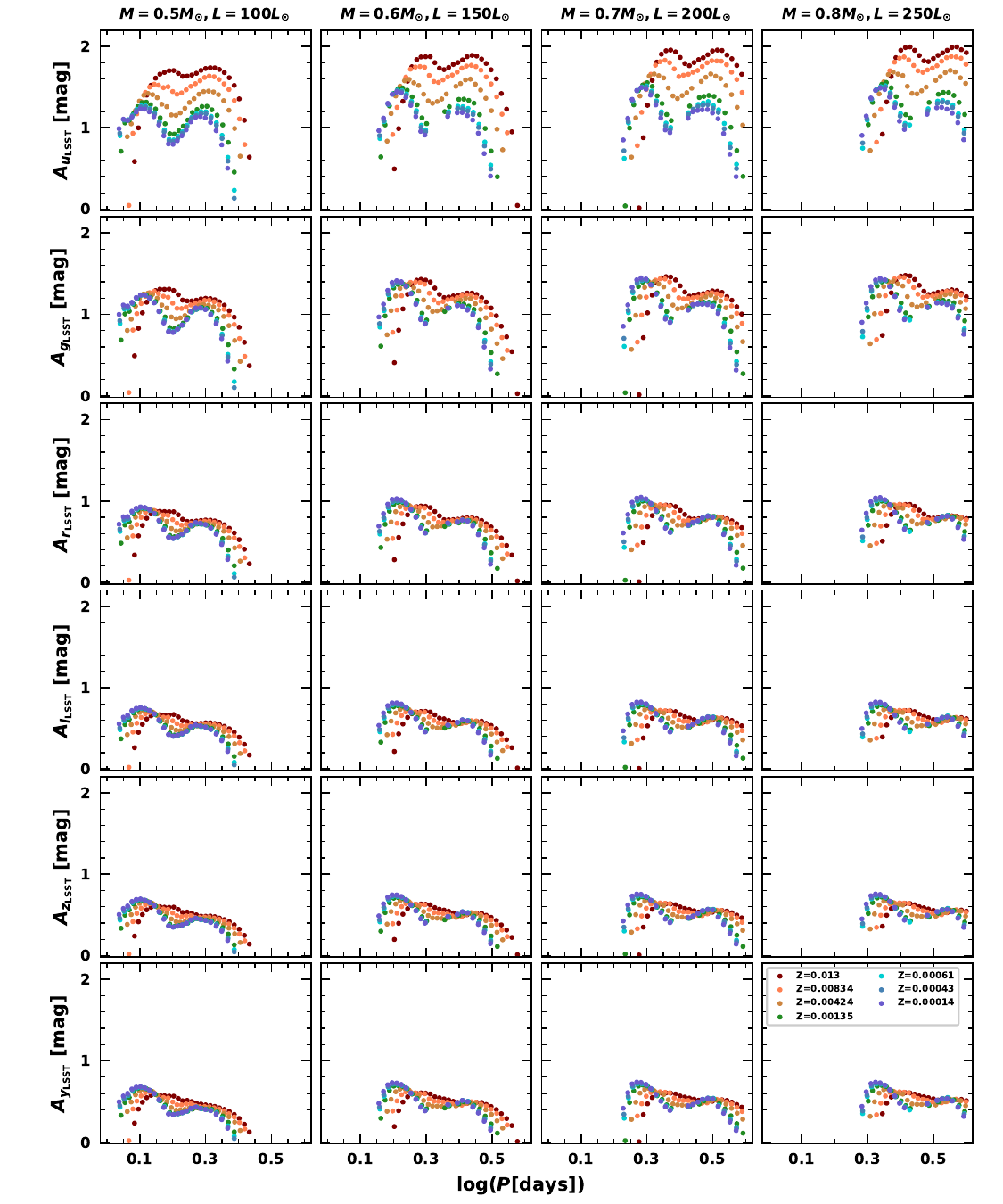}
\caption{The Bailey diagram for the BL~Her models computed using convection parameter set~A as a function of stellar mass, chemical composition and wavelength.}
\label{fig:bailey}
\end{figure*}

The Bailey diagram for the BL~Her models computed using convection parameter set~A as a function of stellar mass, chemical composition and wavelength is displayed in Fig.~\ref{fig:bailey}. Similar results hold for the other convection sets.

\section{Theoretical $PL$ slopes as a function of chemical composition}
\begin{figure*}
\centering
\includegraphics[scale = 0.95]{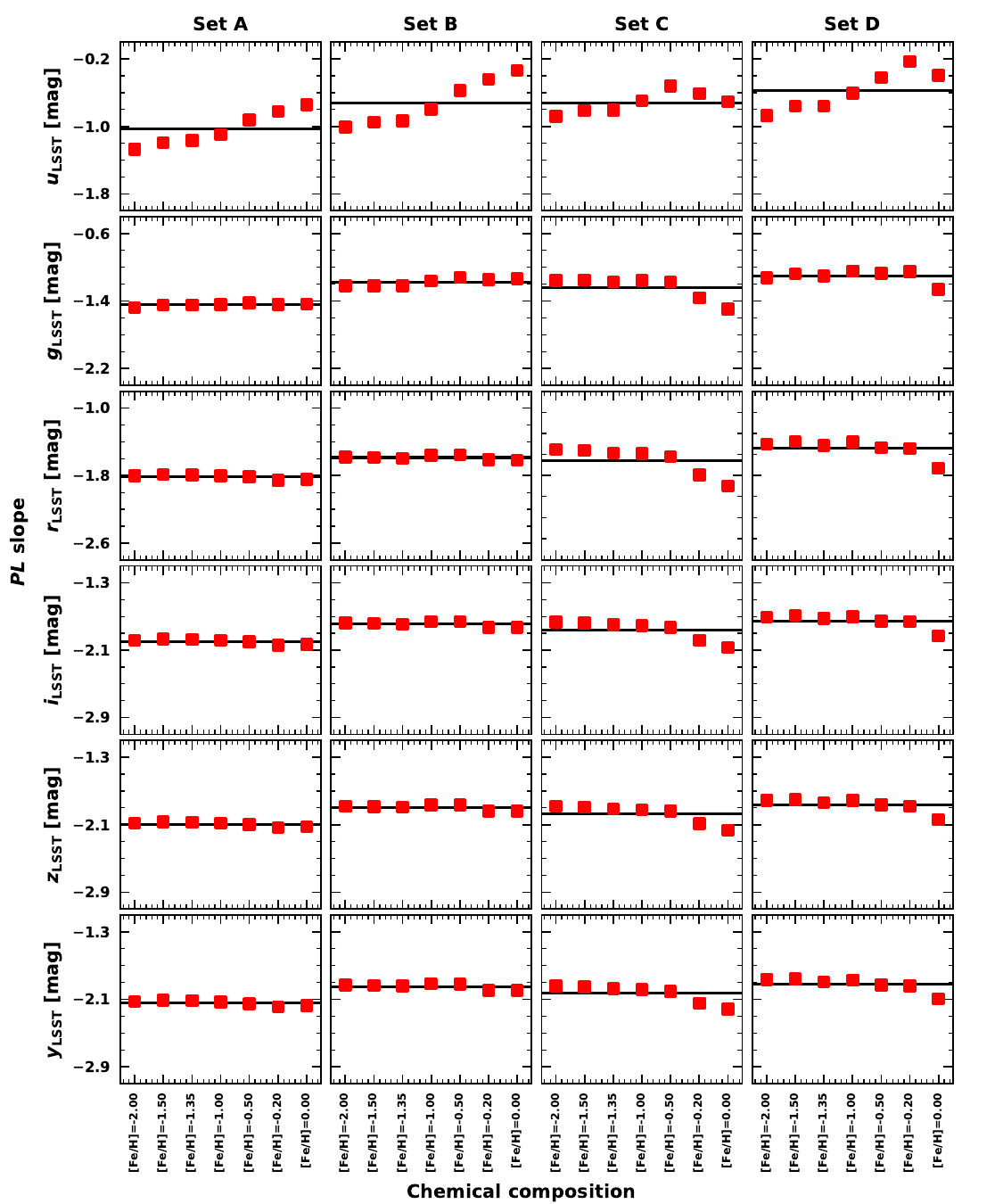}
\caption{Comparison of the $PL$ slopes of the BL~Her models as a function of increasing metallicity (see Table~\ref{tab:composition} for equivalent $ZX$ values) across the different Rubin-LSST filters for the convective parameter sets~A, B, C, and D. The horizontal lines represent the mean values of the slopes in the individual sub-plots. The y-scale is same (2 units) in each panel for a relative comparison.}
\label{fig:PL_diffZ}
\end{figure*}

A comparison of the $PL$ slopes of the BL~Her models as a function of chemical composition across different Rubin-LSST filters for the four sets of convective parameters is  presented in Fig.~\ref{fig:PL_diffZ}.

\section{Modified $PLZ$ and $PWZ$ relations}
The modified $PLZ$ and $PWZ$ relations are presented in Tables~\ref{tab:PLZ_modified} and \ref{tab:PWZ_modified}, respectively.

\begin{table}
\caption{Modified $PLZ$ relations for BL~Her models of the mathematical form $M_\lambda=\alpha+(\beta_1+\beta_2[Fe/H])\log(P)+\gamma\mathrm{[Fe/H]}$ for different wavelengths using different convective parameter sets.}
\centering
\scalebox{0.7}{
\begin{tabular}{c c c c c c c}
\hline\hline
Band & $\alpha$ & $\beta_1$ & $\beta_2$ & $\gamma$ & $\sigma$ & $N$\\
\hline \hline
\multicolumn{7}{c}{Complete set of models ($0.5-0.8M_{\odot}$)}\\
\hline
\multicolumn{7}{c}{Convection set A}\\
\hline
$u$ & 0.969 $\pm$ 0.023 & -0.715 $\pm$ 0.064 & 0.32 $\pm$ 0.059 & 0.125 $\pm$ 0.02 & 0.358 & 3266\\
$g$ & -0.07 $\pm$ 0.02 & -1.416 $\pm$ 0.056 & 0.029 $\pm$ 0.051 & 0.016 $\pm$ 0.018 & 0.311 & 3266\\
$r$ & -0.244 $\pm$ 0.017 & -1.839 $\pm$ 0.045 & -0.027 $\pm$ 0.042 & -0.011 $\pm$ 0.014 & 0.253 & 3266\\
$i$ & -0.246 $\pm$ 0.015 & -2.025 $\pm$ 0.04 & -0.03 $\pm$ 0.037 & -0.005 $\pm$ 0.013 & 0.225 & 3266\\
$z$ & -0.215 $\pm$ 0.014 & -2.119 $\pm$ 0.038 & -0.028 $\pm$ 0.035 & 0.004 $\pm$ 0.012 & 0.214 & 3266\\
$y$ & -0.21 $\pm$ 0.014 & -2.171 $\pm$ 0.037 & -0.031 $\pm$ 0.034 & 0.002 $\pm$ 0.012 & 0.209 & 3266\\
\hline
\multicolumn{7}{c}{Convection set B}\\
\hline
$u$ & 0.97 $\pm$ 0.023 & -0.351 $\pm$ 0.068 & 0.384 $\pm$ 0.061 & 0.11 $\pm$ 0.02 & 0.315 & 2260\\
$g$ & -0.076 $\pm$ 0.021 & -1.122 $\pm$ 0.06 & 0.055 $\pm$ 0.053 & -0.001 $\pm$ 0.018 & 0.277 & 2260\\
$r$ & -0.256 $\pm$ 0.017 & -1.604 $\pm$ 0.05 & -0.018 $\pm$ 0.044 & -0.023 $\pm$ 0.015 & 0.23 & 2260\\
$i$ & -0.261 $\pm$ 0.015 & -1.818 $\pm$ 0.045 & -0.028 $\pm$ 0.04 & -0.014 $\pm$ 0.013 & 0.208 & 2260\\
$z$ & -0.231 $\pm$ 0.015 & -1.923 $\pm$ 0.043 & -0.03 $\pm$ 0.038 & -0.005 $\pm$ 0.013 & 0.199 & 2260\\
$y$ & -0.227 $\pm$ 0.015 & -1.981 $\pm$ 0.042 & -0.034 $\pm$ 0.038 & -0.006 $\pm$ 0.012 & 0.195 & 2260\\
\hline
\multicolumn{7}{c}{Convection set C}\\
\hline
$u$ & 1.173 $\pm$ 0.026 & -0.549 $\pm$ 0.071 & 0.173 $\pm$ 0.067 & 0.203 $\pm$ 0.023 & 0.381 & 2632\\
$g$ & 0.098 $\pm$ 0.022 & -1.38 $\pm$ 0.059 & -0.151 $\pm$ 0.056 & 0.07 $\pm$ 0.019 & 0.318 & 2632\\
$r$ & -0.127 $\pm$ 0.018 & -1.82 $\pm$ 0.048 & -0.171 $\pm$ 0.046 & 0.027 $\pm$ 0.016 & 0.258 & 2632\\
$i$ & -0.155 $\pm$ 0.016 & -2.005 $\pm$ 0.043 & -0.149 $\pm$ 0.041 & 0.027 $\pm$ 0.014 & 0.231 & 2632\\
$z$ & -0.135 $\pm$ 0.015 & -2.104 $\pm$ 0.041 & -0.139 $\pm$ 0.039 & 0.034 $\pm$ 0.013 & 0.219 & 2632\\
$y$ & -0.134 $\pm$ 0.015 & -2.16 $\pm$ 0.039 & -0.136 $\pm$ 0.038 & 0.03 $\pm$ 0.013 & 0.213 & 2632\\
\hline
\multicolumn{7}{c}{Convection set D}\\
\hline
$u$ & 1.12 $\pm$ 0.028 & -0.26 $\pm$ 0.076 & 0.331 $\pm$ 0.068 & 0.156 $\pm$ 0.023 & 0.358 & 2122\\
$g$ & 0.056 $\pm$ 0.023 & -1.137 $\pm$ 0.064 & -0.033 $\pm$ 0.057 & 0.033 $\pm$ 0.02 & 0.302 & 2122\\
$r$ & -0.158 $\pm$ 0.019 & -1.631 $\pm$ 0.053 & -0.094 $\pm$ 0.047 & 0.003 $\pm$ 0.016 & 0.248 & 2122\\
$i$ & -0.18 $\pm$ 0.017 & -1.844 $\pm$ 0.048 & -0.094 $\pm$ 0.042 & 0.009 $\pm$ 0.015 & 0.223 & 2122\\
$z$ & -0.157 $\pm$ 0.016 & -1.954 $\pm$ 0.045 & -0.094 $\pm$ 0.04 & 0.019 $\pm$ 0.014 & 0.212 & 2122\\
$y$ & -0.155 $\pm$ 0.016 & -2.015 $\pm$ 0.044 & -0.097 $\pm$ 0.039 & 0.017 $\pm$ 0.014 & 0.207 & 2122\\
\hline
\multicolumn{7}{c}{Low-mass models only ($0.5-0.6M_{\odot}$)}\\
\hline
\multicolumn{7}{c}{Convection set A}\\
\hline
$u$ & 0.998 $\pm$ 0.036 & -0.129 $\pm$ 0.103 & 0.317 $\pm$ 0.094 & 0.113 $\pm$ 0.031 & 0.321 & 1050\\
$g$ & -0.022 $\pm$ 0.031 & -0.95 $\pm$ 0.087 & -0.029 $\pm$ 0.08 & 0.015 $\pm$ 0.027 & 0.274 & 1050\\
$r$ & -0.172 $\pm$ 0.024 & -1.478 $\pm$ 0.068 & -0.088 $\pm$ 0.062 & -0.011 $\pm$ 0.021 & 0.212 & 1050\\
$i$ & -0.162 $\pm$ 0.02 & -1.713 $\pm$ 0.058 & -0.086 $\pm$ 0.053 & -0.005 $\pm$ 0.018 & 0.181 & 1050\\
$z$ & -0.127 $\pm$ 0.019 & -1.826 $\pm$ 0.054 & -0.082 $\pm$ 0.05 & 0.003 $\pm$ 0.016 & 0.169 & 1050\\
$y$ & -0.121 $\pm$ 0.018 & -1.886 $\pm$ 0.052 & -0.084 $\pm$ 0.048 & 0.002 $\pm$ 0.016 & 0.163 & 1050\\
\hline
\multicolumn{7}{c}{Convection set B}\\
\hline
$u$ & 1.071 $\pm$ 0.032 & 0.071 $\pm$ 0.093 & 0.406 $\pm$ 0.084 & 0.109 $\pm$ 0.028 & 0.254 & 707\\
$g$ & 0.025 $\pm$ 0.027 & -0.766 $\pm$ 0.079 & 0.042 $\pm$ 0.072 & -0.001 $\pm$ 0.024 & 0.217 & 707\\
$r$ & -0.143 $\pm$ 0.021 & -1.322 $\pm$ 0.062 & -0.036 $\pm$ 0.057 & -0.024 $\pm$ 0.019 & 0.17 & 707\\
$i$ & -0.142 $\pm$ 0.018 & -1.573 $\pm$ 0.054 & -0.045 $\pm$ 0.049 & -0.015 $\pm$ 0.016 & 0.147 & 707\\
$z$ & -0.11 $\pm$ 0.017 & -1.691 $\pm$ 0.05 & -0.046 $\pm$ 0.046 & -0.006 $\pm$ 0.015 & 0.138 & 707\\
$y$ & -0.106 $\pm$ 0.017 & -1.754 $\pm$ 0.049 & -0.05 $\pm$ 0.045 & -0.007 $\pm$ 0.015 & 0.134 & 707\\
\hline
\multicolumn{7}{c}{Convection set C}\\
\hline
$u$ & 1.388 $\pm$ 0.041 & -0.444 $\pm$ 0.112 & 0.011 $\pm$ 0.107 & 0.259 $\pm$ 0.036 & 0.354 & 856\\
$g$ & 0.297 $\pm$ 0.034 & -1.311 $\pm$ 0.091 & -0.315 $\pm$ 0.087 & 0.115 $\pm$ 0.029 & 0.288 & 856\\
$r$ & 0.058 $\pm$ 0.026 & -1.764 $\pm$ 0.071 & -0.304 $\pm$ 0.067 & 0.06 $\pm$ 0.023 & 0.222 & 856\\
$i$ & 0.025 $\pm$ 0.022 & -1.953 $\pm$ 0.061 & -0.265 $\pm$ 0.057 & 0.054 $\pm$ 0.02 & 0.191 & 856\\
$z$ & 0.042 $\pm$ 0.021 & -2.054 $\pm$ 0.056 & -0.248 $\pm$ 0.053 & 0.058 $\pm$ 0.018 & 0.177 & 856\\
$y$ & 0.042 $\pm$ 0.02 & -2.111 $\pm$ 0.054 & -0.241 $\pm$ 0.051 & 0.053 $\pm$ 0.017 & 0.17 & 856\\
\hline
\multicolumn{7}{c}{Convection set D}\\
\hline
$u$ & 1.33 $\pm$ 0.04 & -0.199 $\pm$ 0.112 & 0.328 $\pm$ 0.103 & 0.172 $\pm$ 0.036 & 0.335 & 711\\
$g$ & 0.236 $\pm$ 0.033 & -1.058 $\pm$ 0.091 & -0.019 $\pm$ 0.084 & 0.031 $\pm$ 0.029 & 0.274 & 711\\
$r$ & 0.01 $\pm$ 0.026 & -1.556 $\pm$ 0.071 & -0.074 $\pm$ 0.065 & -0.005 $\pm$ 0.023 & 0.212 & 711\\
$i$ & -0.015 $\pm$ 0.022 & -1.773 $\pm$ 0.061 & -0.073 $\pm$ 0.056 & 0.001 $\pm$ 0.02 & 0.183 & 711\\
$z$ & 0.005 $\pm$ 0.02 & -1.883 $\pm$ 0.057 & -0.073 $\pm$ 0.052 & 0.009 $\pm$ 0.018 & 0.17 & 711\\
$y$ & 0.005 $\pm$ 0.02 & -1.944 $\pm$ 0.055 & -0.075 $\pm$ 0.051 & 0.007 $\pm$ 0.018 & 0.165 & 711\\
\hline
\end{tabular}}
\label{tab:PLZ_modified}
\end{table}

\begin{table}
\caption{Modified $PWZ$ relations for BL~Her models of the mathematical form $W_{\lambda_2, \lambda_1}=\alpha+(\beta_1+\beta_2[Fe/H])\log(P)+\gamma\mathrm{[Fe/H]}$ for different wavelengths using different convective parameter sets.}
\centering
\scalebox{0.7}{
\begin{tabular}{c c c c c c c}
\hline\hline
Band & $\alpha$ & $\beta_1$ & $\beta_2$ & $\gamma$ & $\sigma$ & $N$\\
\hline \hline
\multicolumn{7}{c}{Complete set of models ($0.5-0.8M_{\odot}$)}\\
\hline
\multicolumn{7}{c}{Convection set A}\\
\hline
$W_{ug}$ & -3.29 $\pm$ 0.015 & -3.589 $\pm$ 0.04 & -0.875 $\pm$ 0.037 & -0.32 $\pm$ 0.013 & 0.226 & 3266\\
$W_{ur}$ & -1.77 $\pm$ 0.011 & -3.253 $\pm$ 0.029 & -0.464 $\pm$ 0.027 & -0.182 $\pm$ 0.009 & 0.163 & 3266\\
$W_{gr}$ & -0.73 $\pm$ 0.009 & -3.022 $\pm$ 0.025 & -0.183 $\pm$ 0.023 & -0.088 $\pm$ 0.008 & 0.141 & 3266\\
$W_{gi}$ & -0.472 $\pm$ 0.01 & -2.809 $\pm$ 0.026 & -0.104 $\pm$ 0.024 & -0.033 $\pm$ 0.008 & 0.145 & 3266\\
$W_{iz}$ & -0.117 $\pm$ 0.012 & -2.419 $\pm$ 0.032 & -0.023 $\pm$ 0.03 & 0.031 $\pm$ 0.01 & 0.18 & 3266\\
$W_{gy}$ & -0.289 $\pm$ 0.011 & -2.593 $\pm$ 0.029 & -0.064 $\pm$ 0.027 & -0.006 $\pm$ 0.009 & 0.163 & 3266\\
\hline
\multicolumn{7}{c}{Convection set B}\\
\hline
$W_{ug}$ & -3.317 $\pm$ 0.016 & -3.513 $\pm$ 0.047 & -0.964 $\pm$ 0.042 & -0.344 $\pm$ 0.014 & 0.217 & 2260\\
$W_{ur}$ & -1.798 $\pm$ 0.012 & -3.181 $\pm$ 0.035 & -0.524 $\pm$ 0.031 & -0.192 $\pm$ 0.01 & 0.162 & 2260\\
$W_{gr}$ & -0.76 $\pm$ 0.011 & -2.953 $\pm$ 0.031 & -0.223 $\pm$ 0.027 & -0.088 $\pm$ 0.009 & 0.142 & 2260\\
$W_{gi}$ & -0.499 $\pm$ 0.011 & -2.714 $\pm$ 0.032 & -0.134 $\pm$ 0.028 & -0.032 $\pm$ 0.009 & 0.146 & 2260\\
$W_{iz}$ & -0.137 $\pm$ 0.013 & -2.26 $\pm$ 0.038 & -0.037 $\pm$ 0.033 & 0.027 $\pm$ 0.011 & 0.173 & 2260\\
$W_{gy}$ & -0.311 $\pm$ 0.012 & -2.461 $\pm$ 0.035 & -0.085 $\pm$ 0.031 & -0.009 $\pm$ 0.01 & 0.16 & 2260\\
\hline
\multicolumn{7}{c}{Convection set C}\\
\hline
$W_{ug}$ & -3.232 $\pm$ 0.015 & -3.956 $\pm$ 0.04 & -1.157 $\pm$ 0.038 & -0.342 $\pm$ 0.013 & 0.215 & 2632\\
$W_{ur}$ & -1.763 $\pm$ 0.011 & -3.418 $\pm$ 0.03 & -0.602 $\pm$ 0.028 & -0.194 $\pm$ 0.01 & 0.16 & 2632\\
$W_{gr}$ & -0.758 $\pm$ 0.01 & -3.049 $\pm$ 0.026 & -0.224 $\pm$ 0.025 & -0.092 $\pm$ 0.009 & 0.141 & 2632\\
$W_{gi}$ & -0.481 $\pm$ 0.01 & -2.809 $\pm$ 0.027 & -0.146 $\pm$ 0.026 & -0.028 $\pm$ 0.009 & 0.148 & 2632\\
$W_{iz}$ & -0.068 $\pm$ 0.013 & -2.42 $\pm$ 0.034 & -0.107 $\pm$ 0.032 & 0.054 $\pm$ 0.011 & 0.183 & 2632\\
$W_{gy}$ & -0.264 $\pm$ 0.011 & -2.596 $\pm$ 0.031 & -0.128 $\pm$ 0.029 & 0.008 $\pm$ 0.01 & 0.165 & 2632\\
\hline
\multicolumn{7}{c}{Convection set D}\\
\hline
$W_{ug}$ & -3.244 $\pm$ 0.017 & -3.859 $\pm$ 0.047 & -1.162 $\pm$ 0.042 & -0.351 $\pm$ 0.014 & 0.218 & 2122\\
$W_{ur}$ & -1.767 $\pm$ 0.013 & -3.357 $\pm$ 0.035 & -0.629 $\pm$ 0.031 & -0.191 $\pm$ 0.011 & 0.166 & 2122\\
$W_{gr}$ & -0.757 $\pm$ 0.011 & -3.012 $\pm$ 0.031 & -0.264 $\pm$ 0.028 & -0.081 $\pm$ 0.01 & 0.146 & 2122\\
$W_{gi}$ & -0.484 $\pm$ 0.012 & -2.753 $\pm$ 0.032 & -0.172 $\pm$ 0.029 & -0.021 $\pm$ 0.01 & 0.151 & 2122\\
$W_{iz}$ & -0.083 $\pm$ 0.014 & -2.306 $\pm$ 0.039 & -0.095 $\pm$ 0.034 & 0.049 $\pm$ 0.012 & 0.181 & 2122\\
$W_{gy}$ & -0.273 $\pm$ 0.013 & -2.506 $\pm$ 0.035 & -0.132 $\pm$ 0.032 & 0.008 $\pm$ 0.011 & 0.166 & 2122\\
\hline
\multicolumn{7}{c}{Low-mass models only ($0.5-0.6M_{\odot}$)}\\
\hline
\multicolumn{7}{c}{Convection set A}\\
\hline
$W_{ug}$ & -3.182 $\pm$ 0.022 & -3.495 $\pm$ 0.061 & -1.105 $\pm$ 0.056 & -0.29 $\pm$ 0.019 & 0.191 & 1050\\
$W_{ur}$ & -1.643 $\pm$ 0.013 & -3.176 $\pm$ 0.036 & -0.598 $\pm$ 0.034 & -0.167 $\pm$ 0.011 & 0.114 & 1050\\
$W_{gr}$ & -0.591 $\pm$ 0.009 & -2.956 $\pm$ 0.025 & -0.252 $\pm$ 0.023 & -0.082 $\pm$ 0.008 & 0.078 & 1050\\
$W_{gi}$ & -0.343 $\pm$ 0.009 & -2.695 $\pm$ 0.027 & -0.159 $\pm$ 0.024 & -0.03 $\pm$ 0.008 & 0.083 & 1050\\
$W_{iz}$ & -0.016 $\pm$ 0.015 & -2.187 $\pm$ 0.042 & -0.068 $\pm$ 0.038 & 0.03 $\pm$ 0.013 & 0.13 & 1050\\
$W_{gy}$ & -0.177 $\pm$ 0.012 & -2.41 $\pm$ 0.034 & -0.115 $\pm$ 0.032 & -0.006 $\pm$ 0.01 & 0.108 & 1050\\
\hline
\multicolumn{7}{c}{Convection set B}\\
\hline
$W_{ug}$ & -3.216 $\pm$ 0.022 & -3.36 $\pm$ 0.063 & -1.087 $\pm$ 0.058 & -0.34 $\pm$ 0.019 & 0.174 & 707\\
$W_{ur}$ & -1.67 $\pm$ 0.013 & -3.076 $\pm$ 0.039 & -0.592 $\pm$ 0.036 & -0.191 $\pm$ 0.012 & 0.107 & 707\\
$W_{gr}$ & -0.614 $\pm$ 0.01 & -2.879 $\pm$ 0.028 & -0.254 $\pm$ 0.025 & -0.089 $\pm$ 0.008 & 0.077 & 707\\
$W_{gi}$ & -0.356 $\pm$ 0.01 & -2.611 $\pm$ 0.029 & -0.156 $\pm$ 0.026 & -0.034 $\pm$ 0.009 & 0.078 & 707\\
$W_{iz}$ & -0.01 $\pm$ 0.014 & -2.072 $\pm$ 0.04 & -0.049 $\pm$ 0.037 & 0.024 $\pm$ 0.012 & 0.11 & 707\\
$W_{gy}$ & -0.179 $\pm$ 0.012 & -2.307 $\pm$ 0.034 & -0.102 $\pm$ 0.031 & -0.011 $\pm$ 0.01 & 0.095 & 707\\
\hline
\multicolumn{7}{c}{Convection set C}\\
\hline
$W_{ug}$ & -3.087 $\pm$ 0.022 & -3.998 $\pm$ 0.059 & -1.326 $\pm$ 0.056 & -0.331 $\pm$ 0.019 & 0.187 & 856\\
$W_{ur}$ & -1.614 $\pm$ 0.013 & -3.423 $\pm$ 0.036 & -0.7 $\pm$ 0.034 & -0.19 $\pm$ 0.012 & 0.113 & 856\\
$W_{gr}$ & -0.608 $\pm$ 0.009 & -3.029 $\pm$ 0.025 & -0.272 $\pm$ 0.023 & -0.094 $\pm$ 0.008 & 0.078 & 856\\
$W_{gi}$ & -0.325 $\pm$ 0.01 & -2.78 $\pm$ 0.027 & -0.201 $\pm$ 0.026 & -0.025 $\pm$ 0.009 & 0.085 & 856\\
$W_{iz}$ & 0.098 $\pm$ 0.016 & -2.375 $\pm$ 0.042 & -0.191 $\pm$ 0.04 & 0.07 $\pm$ 0.014 & 0.133 & 856\\
$W_{gy}$ & -0.101 $\pm$ 0.013 & -2.559 $\pm$ 0.035 & -0.2 $\pm$ 0.033 & 0.019 $\pm$ 0.011 & 0.111 & 856\\
\hline
\multicolumn{7}{c}{Convection set D}\\
\hline
$W_{ug}$ & -3.156 $\pm$ 0.023 & -3.723 $\pm$ 0.064 & -1.094 $\pm$ 0.059 & -0.407 $\pm$ 0.021 & 0.191 & 711\\
$W_{ur}$ & -1.65 $\pm$ 0.014 & -3.262 $\pm$ 0.04 & -0.58 $\pm$ 0.036 & -0.227 $\pm$ 0.013 & 0.119 & 711\\
$W_{gr}$ & -0.62 $\pm$ 0.01 & -2.946 $\pm$ 0.028 & -0.228 $\pm$ 0.026 & -0.104 $\pm$ 0.009 & 0.083 & 711\\
$W_{gi}$ & -0.339 $\pm$ 0.011 & -2.692 $\pm$ 0.029 & -0.143 $\pm$ 0.027 & -0.039 $\pm$ 0.009 & 0.088 & 711\\
$W_{iz}$ & 0.071 $\pm$ 0.016 & -2.238 $\pm$ 0.044 & -0.073 $\pm$ 0.04 & 0.038 $\pm$ 0.014 & 0.131 & 711\\
$W_{gy}$ & -0.124 $\pm$ 0.013 & -2.44 $\pm$ 0.037 & -0.107 $\pm$ 0.034 & -0.006 $\pm$ 0.012 & 0.111 & 711\\
\hline
\end{tabular}}
\label{tab:PWZ_modified}
\end{table}

\end{appendix}

\end{document}